\documentclass[aps,prb,reprint,groupedaddress,superscriptaddress]{revtex4-2}
\usepackage{amsmath,amssymb,amsfonts,mathrsfs}
\usepackage{graphicx}
\usepackage[matha,mathx]{mathabx}
\usepackage{empheq}
\usepackage{dcolumn}
\usepackage{bm}
\usepackage{hyperref}
\hypersetup{
	colorlinks = true,
	allcolors = blue
}
\usepackage{empheq}
\usepackage{float}
\usepackage{xcolor}
\usepackage[normalem]{ulem}
\allowdisplaybreaks

\renewcommand\rho{\varrho}
\newcommand*\ael{\overline{\alpha}}
\newcommand*\F{\mathcal{F}}
\newcommand*\Fb{\overline{\F}}
\newcommand*\Pc{\mathcal{P}}
\newcommand*\dt{\partial_t}
\newcommand*\du{\partial_u}
\newcommand*\duu{\partial_{uu}}

\newcommand*\eps{\varepsilon}
\newcommand*\ka{\overline{\kappa}}
\newcommand*\kb{\text{k}_{\scriptscriptstyle{\rm B}}}
\newcommand*\Temp{\mathsf{T}}
\newcommand*\req{\rho{\scriptscriptstyle{\rm eq}}}

\newcommand*\rnb{\ol{\rho}_n}
\newcommand*\rnh{\hat{\rho}_n}
\newcommand*\rnt{\tilde{\rho}_n}
\newcommand*\LO{L_0}
\newcommand*\LL{L}
\newcommand*\bigO{\mathcal{O}}
\newcommand*\ud{\,\mathrm{d}}
\newcommand*\myPadding{-0.75em}

\renewcommand{\ol}[1]{\overline{#1}}
\renewcommand\Re{\operatorname{Re}}

\begin{document}


\title{Scaling laws for step bunching on vicinal surfaces: \\ the role of the dynamical and chemical effects}

\author{L. Benoit-\phantom{}-Maréchal}
\affiliation{LMS, \'{E}cole polytechnique, CNRS, Institut polytechnique de Paris, 91128 Palaiseau, France}
\affiliation{LPICM, \'{E}cole polytechnique, CNRS, Institut polytechnique de Paris, 91128 Palaiseau, France}
\author{M. E. Jabbour}
\affiliation{LMS, \'{E}cole polytechnique, CNRS, Institut polytechnique de Paris, 91128 Palaiseau, France}
\affiliation{D\'{e}partement de M\'{e}canique, \'{E}cole polytechnique, 91128 Palaiseau, France}
\author{N. Triantafyllidis}
\affiliation{LMS, \'{E}cole polytechnique, CNRS, Institut polytechnique de Paris, 91128 Palaiseau, France}
\affiliation{D\'{e}partement de M\'{e}canique, \'{E}cole polytechnique, 91128 Palaiseau, France}
\affiliation{Aerospace Engineering Department \& Mechanical Engineering Department (emeritus),\\
The University of Michigan, Ann Arbor, MI 48109-2140, USA}

\date{\today}

\begin{abstract}
We study the evolution of step bunches on vicinal surfaces using a thermodynamically consistent step-flow model that (i) circumvents the quasistatic approximation that prevails in the literature by accounting for the dynamics of adatom diffusion on terraces and attachment-detachment at steps (referred to collectively as the \emph{dynamical effect}), and (ii) generalizes the expression of the step chemical potential by incorporating the necessary coupling between the diffusion fields on adjacent terraces (referred to as the \emph{chemical effect}). Having previously shown that these dynamical and chemical effects can explain the onset of step bunching without recourse to the inverse Ehrlich-Schwoebel (iES) barrier or other extraneous mechanisms, we are here interested in the evolution of step bunches beyond the linear-stability regime. In particular, the numerical resolution of the step-flow free boundary problem yields a robust power-law coarsening of the surface profile, with the bunch height growing in time as $H \sim t^{1/2}$ and the minimal interstep distance as a function of the number of steps in the bunch cell obeying $\ell_{min} \sim N^{-2/3}$. Although these exponents have previously been reported, this is the first time such scaling laws are obtained in the absence of an iES barrier or adatom electromigration. In order to validate our simulations, we take the continuum limit of the discrete step-flow system via Taylor expansions with respect to the terrace size, leading to a novel nonlinear evolution equation for the surface height. We investigate the existence of self-similar solutions of this equation and confirm the 1/2 coarsening exponent obtained numerically for $H$. We highlight the influence of the combined dynamical-chemical effect and show that it can be interpreted as an effective iES barrier in the setting of the standard Burton-Cabrera-Frank theory. Finally, we use a Padé approximant to derive an analytical expression for the velocity of steadily moving step bunches and compare it to numerical simulations.

\end{abstract}

\maketitle


\section{Introduction}
Step bunching is a morphological instability on vicinal surfaces whereby straight atomic steps deviate from an equidistant configuration and coalesce, resulting in an alternating pattern of step bunches and wide flat terraces. The study of the characteristic length scales of these patterns which coarsen in time is fundamental to our understanding of the microscopic mechanisms governing crystal growth in the step-flow regime, thus paving the way for such applications as the nanopatterning of semiconductor surfaces \cite{Ronda2004, Wise2005}.

Several mechanisms have been proposed to explain the observed bunching of steps on various semiconductor and metallic surfaces. They include the inverse Ehrlich-Schwoebel (iES) barrier, i.e., the preferential attachment of terrace adatoms to descending steps \cite{Krug2005}; the anisotropy of adatom diffusion on the terraces of reconstructed surfaces \cite{Myslivecek2002}; adatom electromigration when the substrate is heated by an electric current \cite{Latyshev1989,Stoyanov1991}; the presence of impurities, real or effective, that hinder step motion \cite{Kandel1994}; chemical reactions between different species during growth of multicomponent crystals, resulting in an effective iES barrier \cite{Pimpinelli2000}; and edge diffusion \cite{Pierre-Louis1999,Murty1999}. Recently, Guin et al. \cite{Guin2020} revisited the bunching instability by means of a step-flow model derived from the thermodynamics of nonequilibrium processes \cite{Cermelli2005}. By accounting for the dynamics of both adatom diffusion on terraces and their attachment to and detachment from steps (collectively referred to as the \emph{dynamical effect}), this model goes beyond the quasistatic approximation that prevails in the literature on step instabilities. Further, in this model, the expression of the step chemical potential that derives from the kinetic relation linking the driving force at a given step to its velocity generalizes the one found in the literature by accounting for the necessary contribution of the adjacent terraces in the form of the jump in the adatom grand canonical potential. The resulting coupling between the diffusion fields on adjacent terraces is referred to as the \emph{chemical effect}. Importantly, these dynamical and chemical effects are unaccounted for in the stability analyses that take the standard Burton-Cabrera-Frank (BCF) theory as their starting point. However, there is no \emph{a priori} justification for the neglect of these effects. Indeed, as shown in \cite{Guin2020,Guin2021A,*Guin2021B}, their combination significantly alters the stability analysis of step dynamics, even in the low-deposition regime where the quasistatic approximation is assumed to hold, and can quantitatively explain the onset of step bunching on such surfaces as Si(111), where uncertainty remains about the existence of an iES barrier \cite{Voigtlander1995,Ichimiya1996,Chung2002,Rogilo2013}, and GaAs(001), where a direct ES barrier is believed to exist \cite{Smilauer1995,Krug1997,Salmi1999}.

In this work, we aim to investigate step bunching beyond the linear-stability regime. Specifically, we derive scaling relations for the evolution of step bunches induced by the combined dynamical and chemical effects, and show that they suffice to reproduce the coarsening behavior observed experimentally \cite{Omihomma2005}, circumventing the need for the controversial iES barrier. The remainder of the article proceeds as follows. In Section \ref{sec:model}, we summarize the thermodynamically consistent step-flow model that serves as our starting point, highlighting the terms that are unaccounted for in the standard BCF theory. Section \ref{sec:scaling_laws_numerical} is devoted to the scaling laws for the coarsening process obtained from numerical simulations of step flow. The continuum limit, whereby a continuous function is used to describe the surface height, is derived in Section \ref{sec:continuum_limit} via Taylor expansions with respect to the terrace size. Although the nonlinear evolution equation that governs the surface height has a similar structure to that derived elsewhere, its coefficients are modified by the inclusion of the dynamical and chemical effects, allowing for the bunching instability to develop even in the presence of a direct ES barrier. In Section \ref{sec:scaling_laws_theory}, the exponents that enter the scaling laws are extracted analytically and found to be in agreement with our numerical estimates. Finally, an analytical expression for the velocity of steadily moving step bunches is derived using a Padé approximant of the continuum limit, and compared with simulation results of the discrete step-flow equations.

\section{Model}\label{sec:model}
\subsection{Generalized BCF free-boundary problem}
Our starting point is a thermodynamically consistent generalization of the BCF model that circumvents the quasistatic approximation which prevails in the literature on step dynamics and includes terms that are unaccounted for in the step boundary conditions of the standard model, with important implications on the stability of the vicinal surface with respect to bunching \cite{Guin2020,Guin2021A,*Guin2021B}. 

\begin{figure}[b]
\centering
	\includegraphics[width=0.48\textwidth]{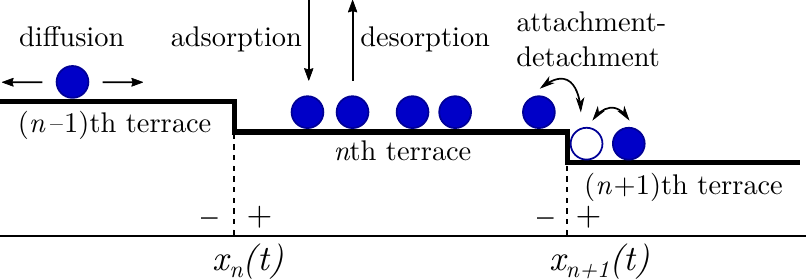}
	\vspace{\myPadding}
	\caption{One-dimensional schematic of successive straight atomic steps, depicting the microscopic mechanisms that govern step-flow growth.}
	\label{fig:schematic_steps}
\end{figure}

In the present one-dimensional setting, straight steps are represented by their positions $\{x_n(t)\}_{n\in\mathbb{N}^\ast}$ along the $x$ axis (Fig.~\ref{fig:schematic_steps}). Let $\rho_n(x,t)$ and $\jmath_n(x,t)$ be the adatom density and diffusive flux on the $n$th terrace $(x_n(t),x_{n+1}(t))$,  and denote by $r_n(x,t)$ the adsorption-desorption flux on the same terrace. Mass balance has the form $\partial_t\rho_n\!=\!-\partial_x\jmath_n\!+\!r_n$. Let $\mu_n(x,t)$ be the adatom chemical potential on the $n$th terrace and denote by $\mu^v$ the chemical potential in the vapor on top of the vicinal surface. Assuming the constitutive relations $\jmath_n\!=\!-\rho_n M\partial_x\mu_n$ and $r_n\!=-\!\sigma(\mu_n\!-\!\mu^v)$, with $M\!>\!0$ the adatom mobility and $\sigma\!>\!0$ the adsorption-desorption coefficient, mass balance on the $n$th terrace takes the form
\begin{equation}
\partial_t\rho_n=-\partial_x(\rho_n M\partial_x\mu_n)-\sigma(\mu_n-\mu^v).
\label{PDE}
\end{equation}

Given a field $\varphi_n(x,t)$ defined on the $n$th terrace, let $\varphi^+_n(t)\!=\!\varphi_n(x_n(t),t)$ be its limiting value at the $n$th step. As the $n$th step is approached from the $n$th terrace, mass balance yields the condition $\rho^+_n\dot{x}_n\!-\!\jmath^+_n\!=\!J^+_n$, where the superimposed dot denotes differentiation with respect to time and $J^+_n(t)$ is the flux of adatoms from the $n$th terrace to the $n$th step. Assuming the constitutive relation $J^+_n\!=\!\gamma_+(\mu^+_n\!-\!\mu^s_n)$, with $\mu^s_n$ the chemical potential of the $n$th step and $\gamma_+\!>\!0$ the attachment-detachment coefficient to a step from its lower terrace, we get
\begin{equation}
\rho^+_n(\dot{x}_n+M(\partial_x\mu_n)^+)=\gamma_+(\mu^+_n-\mu^s_n).
\label{BC1}
\end{equation}

Similarly, let $\varphi^-_n(t)\!=\!\varphi_n(x_{n+1}(t),t)$ be the limiting value of $\varphi_n$ at the $(n\!+\!1)$th step. Approaching the $(n\!+\!1)$th step from the $n$th terrace, mass balance yields the condition $-\rho^-_n\dot{x}_{n+1}\!+\!\jmath^-_n\!=\!J^-_{n+1}$, with $J^-_{n+1}\!=\!\gamma_-(\mu^-_{n+1}\!-\!\mu^s_n)$ the flux of adatoms to the $(n\!+\!1)$th step from its upper terrace and $\gamma_-$ the corresponding attachment-detachment coefficient. Thus, 
\begin{equation}
-\rho^-_n(\dot{x}_{n+1}+M(\partial_x\mu_n)^-)=\gamma_-(\mu^-_n-\mu^s_{n+1}).
\label{BC2}
\end{equation}

We refer to the transient term $\partial_t\rho_n$ in \eqref{PDE} and the \textit{advective} terms $\rho^+_n\dot{x}_n$ and $\rho^-_n\dot{x}_{n+1}$ in \eqref{BC1} and \eqref{BC2} collectively as the \emph{dynamical effect}, in contrast to the quasistatic approximation (in which these terms are absent) that prevails in the literature on step dynamics \cite{Ghez1988,Krug2005,Michely2012}.

Let $a$ and $\psi^c$ be the lattice parameter and free-energy density of the undeformed crystal, and denote by $\omega_n\!=\!\psi_n\!-\!\mu_n\rho_n$ the grand canonical potential of adatoms on the $n$th terrace, with $\psi_n$ the corresponding Helmholtz free-energy density. The step chemical potential satisfies the relation
\begin{equation}
\mu^s_n=\mu^c-a^2([\![\omega]\!]_n+\mathfrak{f}_n+\beta\dot{x}_n),
\label{GT}
\end{equation}
where $\mu^c\!=\!a^2\psi^c$ is the free energy per atom in the bulk, ${[\![\omega]\!]_n\!=\!\omega^+_n\!-\!\omega^-_{n-1}}$ is the jump of $\omega$ across the $n$th step, $\beta$~is the step kinetic coefficient, and $\mathfrak{f}_n$ is the contribution to the driving force at the $n$th step of the elastic fields generated by the other steps on the vicinal surface. For homoepitaxial growth, $\mathfrak{f}_n$ can be approximated by
\begin{equation}
\mathfrak{f}_n=-\alpha\sum_{i\in\mathbb{N}^\ast}\big\{(x_{n+i}-x_n)^{-3}-(x_n-x_{n-i})^{-3}\big\},
\label{f}
\end{equation}
where $\alpha$ depends on the Young modulus and Poisson's ratio of the crystal, and the strength of the dipole-dipole interactions between steps \cite{Stewart1994,Tersoff1995}.
In two space dimensions, an additional term $-\tilde{\psi}^s_n K_n$ appears on the right side of \eqref{GT}, with $K_n$ the curvature of the $n$th step and $\tilde{\psi}^s_n$ its stiffness \cite{Cermelli2005,Michely2012}. Hence, even in the present one-dimensional setting, we refer to \eqref{GT} as the generalized Gibbs--Thomson relation. It states that the chemical potential of the $n$th step differs from its bulk counterpart $\mu^c$, and that the difference consists of three contributions: one, $-a^2[\![\omega]\!]_n$, from the adatoms on the adjacent terraces; another, $a^2\mathfrak{f}_n$, from the elastic bulk; and a third, $a^2\beta\dot{x}_n$, akin to kinetic undercooling in solidification problems \cite{Davis2001}. Hereafter, we neglect this last contribution by assuming that the terraces are in phase equilibrium at the steps, which is tantamount to taking $\beta\!=\!0$. 

We refer to the terrace contribution as the \emph{chemical effect}, since chemical equilibrium between two phases separated by a thermodynamically structureless interface entails the continuity of the grand canonical potential. In the present context, the adjacent terraces can be viewed as distinct phases characterized by different adatom densities and the step separating them as an interface endowed with a free-energy density and chemical potential.

At the $n$th step, mass balance states that the step velocity is proportional to the intake of adatoms from the adjacent terraces
\begin{equation}
\dot{x}_n=a^2(J^+_n+J^-_n),
\label{V}
\end{equation}
a relation that bears resemblance to the Stefan condition, which in the setting of solidification derives from energy balance.

With the constitutive prescription of the Helmholtz free-energy density $\psi_n(\rho_n)$, so that the adatom chemical potential $\mu_n\!=\!\partial\psi_n/\partial\rho_n$ and grand canonical potential $\omega_n$ are given functions of the adatom density, \eqref{PDE}, \eqref{BC1}, \eqref{BC2}, and \eqref{V} form a free-boundary problem whose unknowns are the adatom densities $\{\rho_n\}_{n\in\mathbb{N}^\ast}$ and step positions $\{x_n\}_{n\in\mathbb{N}^\ast}$.

Next, assume that the adatoms behave like an ideal lattice gas. Let $\kb$ be the Boltzmann constant and $\mathsf{T}$ the absolute temperature, and denote by $D\!=\!\kb\Temp M$ the adatom diffusivity. The adatom balance \eqref{PDE} reduces to a nonlinear reaction-diffusion  equation on the $n$th terrace
\begin{equation}
\partial_t\rho_n=D\partial^2_{xx}\rho_n-\sigma\Big\{\kb\Temp\ln\Big(\frac{\rho_n}{\req}\Big)+\mu^c-\mu^v\Big\},
\label{PDEi}
\end{equation}
where $\req$ is the equilibrium adatom density defined by $\mu_n(\req)\!=\!\mu^c$ for a train of equidistant steps at rest. Moreover, the boundary condition \eqref{BC1} at the $n$th step now has the form
\begin{equation}
\rho^+_n\dot{x}_n+D(\partial_x\rho_n)^+=J^+_n,
\label{BC1i}
\end{equation}
where, denoting $\kappa_+\hspace{-0.5em}=\!\kb\Temp\gamma_+/\req$, the flux of adatoms from the $n$th terrace to the $n$th step is given by
\begin{equation}
J^+_n=\kappa_+\Big\{\req\ln\Big(\frac{\rho^+_n}{\req}\Big)-a^2\req\Big([\![\rho]\!]_n+\frac{\mathfrak{f}_n}{\kb\Temp}\Big)\Big\}.
\end{equation}

Similarly, the boundary condition \eqref{BC2} at the $(n\!+\!1)$th step can be rewritten as
\begin{equation}
-\rho^-_n\dot{x}_{n+1}-D(\partial_x\rho_n)^-=J^-_{n+1},
\label{BC2i}
\end{equation}
where, denoting  $\kappa_-\hspace{-0.5em}=\!\kb\Temp\gamma_-/\req$, the flux of adatoms from the $n$th terrace into the $(n\!+\!1)$th step reads
\begin{equation}
J^-_{n+1}=\kappa_-\Big\{\req\ln\Big(\frac{\rho^-_n}{\req}\Big)-a^2\req\Big([\![\rho]\!]_{n+1}+\frac{\mathfrak{f}_{n+1}}{\kb\Temp}\Big)\Big\}.
\end{equation}

Assuming small departures of the adatom density from its equilibrium value, $|\rho_n\!-\!\req|/\req\!\ll\!1$, we can linearize the logarithmic terms in \eqref{V}, \eqref{PDEi}, \eqref{BC1i}, and \eqref{BC2i}. In particular, \eqref{PDEi} reduces to
\begin{equation}
\partial_t\rho_n=D\partial^2_{xx}\rho_n+\F-\nu\rho_n,
\label{RD}
\end{equation}
where $\F\!=\!\sigma(\mu^v\!-\!\mu^c\!+\!\kb\Temp)$ can be viewed as a constant deposition flux and $\nu\!=\!\sigma\kb\Temp/\req$ as a desorption rate. In what follows, we are interested in temperatures that are sufficiently low for desorption to be negligible.

Let $\LO$ be the initial terrace width and denote by $\Pc\!=\!\F a^2\LO/D$ the P\'eclet number. We introduce the dimensionless variables
\begin{equation}\label{eq:normalized_txr}
\ol{t}=\frac{\Pc Dt}{\LO^2},\quad\ol{x}=\frac{x}{\LO},\quad\rnb=\frac{\rho_n}{\req},
\end{equation}
and by making the change of spatial variable
\begin{equation}
u=\frac{\overline{x}-\ol{x}_n(\ol{t})}{\ol{x}_{n+1}(\ol{t})-\ol{x}_n(\ol{t})},
\end{equation}
map the free-boundary problem over the $n$th terrace onto the fixed interval $[0,1]$. Specifically, letting $\Fb\!=\!\F\LO^2/\req D$ denote the scaled deposition flux, \eqref{RD} can be rewritten in dimensionless form
\begin{equation}
\chi_a\Pc\Big\{\partial_{\ol{t}}\rnb-\frac{(\dot{\ol{x}}_n+\dot{\ol{s}}_nu)}{\ol{s}_n}\partial_u\rnb\Big\}=\frac{1}{\ol{s}^2_n}\partial^2_{uu}\rnb+\Fb,
\label{eq:ale}
\end{equation}
where $\ol{s}_n(t)\!=\!\ol{x}_{n+1}(t)\!-\!\ol{x}_n(t)$ is the scaled width of the $n$th terrace and $\chi_a\!=\!1$ is a parameter introduced to track the dynamical effect, in the sense that formally setting $\chi_a\!=\!0$ corresponds to the quasistatic approximation. 

Let $\Theta\!=\!a^2\req$ be the equilibrium adatom coverage, so that $\Pc\!=\!\Fb\Theta$, and introduce the dimensionless parameters
\begin{equation}
\ka=\frac{\kappa_-\LO}{D}\quad\text{and}\quad S=\frac{\kappa_+}{\kappa_-},
\end{equation}
where $\ka$ measures the strength of the step attachment-detachment kinetics relative to the terrace diffusion kinetics and $S$ specifies the nature of the ES effect: $S\!>\!1$ corresponds to the direct ES barrier, $S\!<\!1$ to its inverse, and $S\!=\!1$ to symmetric adatom attachment to and detachment from the steps. We refer to the limit $\ka\!\ll\!1$ as the kinetic-limited (KL) regime and $\ka\!\gg\!1$ as the diffusion-limited (DL) regime. Eq.~\eqref{eq:ale} is supplemented by the boundary conditions
\begin{figure*}
	\begin{minipage}[c]{0.7\textwidth}
		\includegraphics[width=\textwidth]{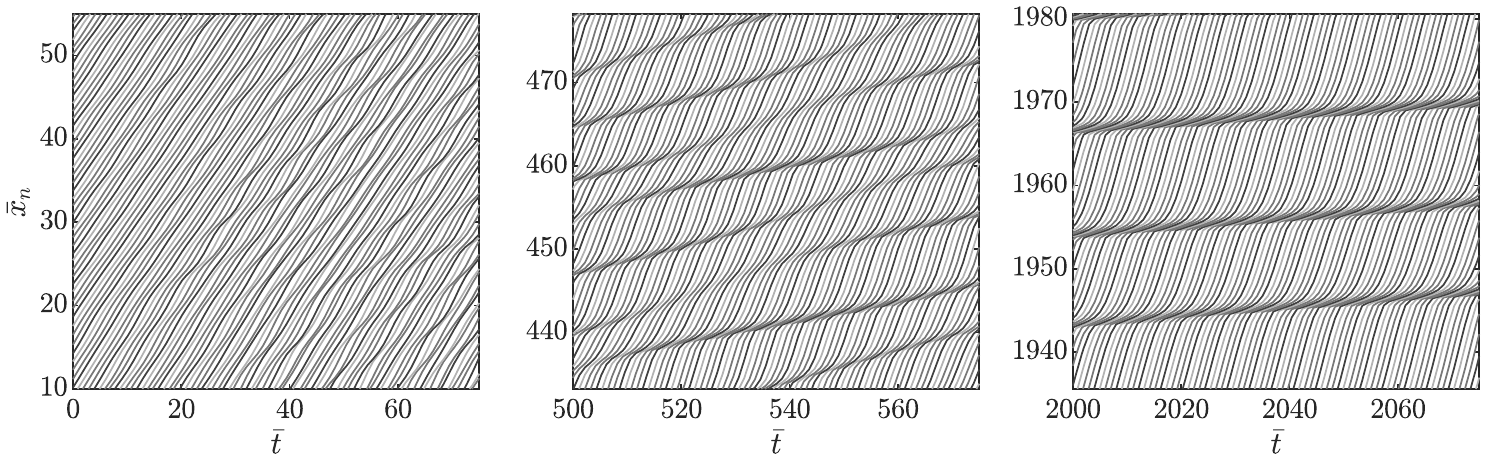}
	\end{minipage}\hfill
	\begin{minipage}[c]{0.27\textwidth}
		\caption{Spatiotemporal evolution of a vicinal surface with 500-step periodicity. Each line is a step trajectory. Only 45 steps are shown for clarity. \protect\\ The chosen parameters are: $S\!=1\!$, $\Theta\!=\!0.02$, $\Fb\!=\!10^{-4}$, $\ka\!=\!10^{-2}$, and $\ael\!=\!10^{-5}$.}\vspace{2em}
		\label{fig:spatiotemporal_diagram}
	\end{minipage}
\end{figure*}
\begin{equation}
\chi_a\ol{s}_n\rnb^+\dot{\ol{x}}_n+(\partial_u\rnb)^+=\ol{s}_n\ol{J}^+_n
\label{BC1d}
\end{equation}
at the $n$th step and 
\begin{equation}
-\chi_a\ol{s}_n\rnb^-\dot{\ol{x}}_{n+1}-(\partial_u\rnb)^-=\ol{s}_n\ol{J}^-_{n+1}
\label{BC2d}
\end{equation}
at the $(n\!+\!1)$th step. In \eqref{BC1d}, the dimensionless adatom flux from the $n$th terrace to the $n$th step is given by
\begin{equation}
\ol{J}^+_n=\ka S(\rnb^+-1-\chi_c\Theta[\![\ol{\rho}]\!]_n+\ol{\mathfrak{f}}_n),
\label{J+d}
\end{equation}
where, letting $\ol{\alpha}\!=\!a^2\alpha/\kb\Temp\LO^3$, the scaled elastic contribution to the driving force at the $n$th step reads
\begin{equation}
\ol{\mathfrak{f}}_n=-\ol{\alpha}\sum_{i\in\mathbb{N}^\ast}\big\{(\ol{x}_{n+i}-\ol{x}_n)^{-3}-(\ol{x}_n-\ol{x}_{n-i})^{-3}\big\},
\label{fd}
\end{equation}
and in \eqref{BC2d} the dimensionless adatom flux from the $n$th terrace to the $(n\!+\!1)$th step has the form 
\begin{equation}
\ol{J}^-_{n+1}=\ka (\rnb^--1-\chi_c\Theta[\![\ol{\rho}]\!]_{n+1}+\ol{\mathfrak{f}}_{n+1}).
\label{J-d}
\end{equation}
The parameter $\chi_c\!=\!1$ is introduced in \eqref{J+d} and \eqref{J-d} in order to track the chemical effect, in the sense that by formally setting $\chi_c\!=\!0$ (in addition to $\chi_a\!=\!0$) the step conditions \eqref{BC1d} and \eqref{BC2d} reduce to those of the standard BCF model \cite{Ghez1988,Krug2005}.

Finally, the adatom balance \eqref{V} at the $n$th step can be rewritten in nondimensional form as
\begin{equation}
\Pc\dot{\ol{x}}_n=\Theta(\ol{J}^+_n+\ol{J}^-_n).
\label{Vd}
\end{equation}

\subsection{Numerical solution method}
Recall that the quasistatic BCF model is recovered by setting $\chi_a\!=\!\chi_c\!=\!0$ in \eqref{eq:ale}, \eqref{BC1d}, \eqref{BC2d} and \eqref{Vd}. This model can easily be solved for the adatom densities, yielding a system of ODE's for the step positions whose numerical integration is straightforward \cite{Stoyanov1994,Frisch2005,Krug2005_scaling}. Since we are interested in the influence of the dynamical and chemical effects on the onset and evolution of step bunching, this method is not suitable for our model.

An alternative to the BCF model that includes dynamical effects is the phase-field approximation, whereby step flow is governed by a system of two coupled PDE's for a global adatom density field and an order parameter (the phase field) which is constant on the terraces but varies rapidly inside narrow transition regions around the steps \cite{Liu1994,Otto2004}. The main feature of the phase-field model is that it automatically captures such topological changes as island formation or step coalescence, making it particularly efficient at predicting the evolution of island shapes in two space dimensions \cite{Torabi2009,Hu2012}. However, in the present one-dimensional setting and in the absence of nucleation and coalescence, the phase-field model is not more advantageous than the direct numerical resolution of the sharp-interface free boundary problem \eqref{eq:ale}, \eqref{BC1d}, \eqref{BC2d} and \eqref{Vd}. Further, since our objective is to obtain scaling laws for the coarsening of step bunches, we need to simulate large numbers of steps in order to mitigate finite-size effects, whereas phase-field simulations are typically limited to small numbers of steps \cite{Ratz2004,Yu2011}.

Thus, in order to simulate the evolution of a sufficient number of steps while retaining the transient and advective terms in the free boundary problem, we use finite elements with a second-order interpolation function to discretize the terraces by the Galerkin method (cf.~Appendix \ref{app:galerkin} for details), and solve for the adatom densities and the step positions concomitantly. For computational efficiency, the number of elements per terrace is reduced to a minimum. Convergence tests show that using only one element already offers a high degree of accuracy. For the same reason, we only consider the first five terms in \eqref{fd}.

To integrate the resulting ODE's, we rely on an implicit scheme since the $1/\ol{s}_n^2$ factor in \eqref{eq:ale} diverges to infinity when bunching occurs ($\ol{s}_n\!\rightarrow\!0$), making the equation extremely stiff and causing explicit solvers to fail. We use Julia's implementation of Sundials' CVODE routine with Backward Differentiation Formula, which implements a variable step, variable order, multistep method \cite{Julia,Sundials}. To improve the solver efficiency, the analytical expression of the jacobian is provided and sparse matrices are used. 

We impose periodic boundary conditions on the space domain and initialize the system in two different configurations. Under \emph{natural bunching conditions}, the integration is initiated from a vicinal surface with 500 steps whose deviation from their equidistant equilibrium position follows a uniform distribution in $[-0.1,0.1]$. This leads to a surface profile consisting of many bunches separated by large terraces (Fig.~\ref{fig:height_profile}), which slowly coarsens. Under \emph{forced bunching conditions}, a number of steps (from 10 to 200) are initially placed in close proximity ($\forall n,\ \ol{s}_n\!=\!0.1$ is arbitrarily chosen) so that, as time progresses, they will relax towards a stable arrangement, providing the actual quasisteady bunch shape. For easier comparison with existing results, we only consider nearest-neighbors elastic interaction in this configuration. The model parameters are selected to trigger the step bunching instability, as discussed in \cite{Guin2020}.

In all the displayed figures, the normalized time $\ol{t}$ defined in \eqref{eq:normalized_txr} is used, which represents the number of deposited monolayers, and distances are normalized by the initial terrace length $\LO$. An example of the formation and evolution of step bunches is shown in a spatiotemporal diagram (Fig.~\ref{fig:spatiotemporal_diagram}), where each line represents a step trajectory. 
The lines are initially straight and parallel as the steps propagate at a constant velocity (obtained from the steady-state solution), until the instability develops (visible already after 20 monolayers) and the lines swerve towards each other as steps start to coalesce. 

At later times, the bunched structure is clearly visible. Note that step bunching is a dynamic process: a bunch is not a fixed entity which contains identifiable steps (as can be the case under certain electromigration conditions \cite{Sato1999,Homma2000,Toktarbaiuly2018}), but rather continually emits and receives steps to and from neighboring bunches (crossing steps). Moreover, the number of bunches decreases as they increase in size: this is a direct manifestation of the coarsening process. Finally, bunches move much slower than steps, with a seemingly inverse correlation between their velocity and size.

\section{Coarsening behaviour}\label{sec:scaling_laws_numerical}

\begin{figure}[t]
\centering
	\includegraphics[width=0.4\textwidth]{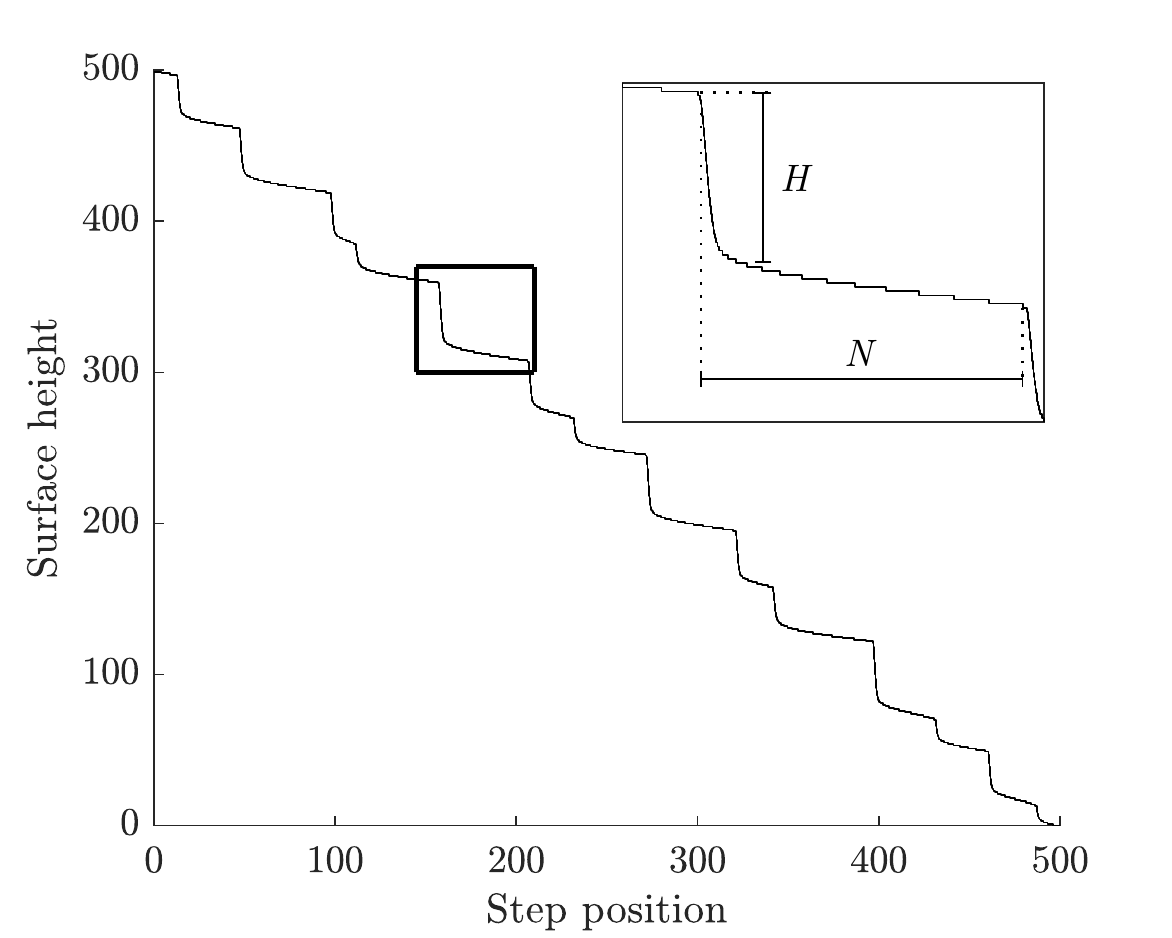} 
	\vspace{\myPadding}
	\caption{Height profile of a surface with 500 steps after deposition of $3 \times 10^4$ monolayers. Same parameters as in Fig. \ref{fig:spatiotemporal_diagram}.}
	\label{fig:height_profile}
\end{figure}

Before proceeding with the quantitative analysis of the coarsening process, we introduce some characteristic parameters to describe the bunched surface. A difference is made between a \emph{bunch}, which corresponds to the high step-density region only, and a \emph{bunch cell}, which comprises a bunch and the terraces running to the next bunch. We denote $H$ the height and $W$ the width of a bunch, $\ell_{min}$ and $\ell_{max}$ the narrowest and widest terrace on the surface, and $N$ the number of steps in a bunch cell, which corresponds to the distance between bunches in units of $\LO$, as shown in Fig.~\ref{fig:height_profile}. Since vicinality requires the average slope of the surface to remain constant, $H \sim N$.

Among these parameters, we choose to focus on the scaling of $H$ with $\ol{t}$ and the scaling of $\ell_{min}$ with $N$, as they are the most reliable indicators and can be easily computed from theoretical models for comparison. Indeed, monitoring the evolution of bunches requires the introduction of an arbitrary threshold on the interstep distance, which determines whether steps belong to the same bunch or not. While some quantities (e.g., $\ell_{min}$ and $\ell_{max}$) are independent of any threshold, others (e.g., $H$ and $W$) are sensitive to this choice, especially so because of the asymmetrical distribution of crossing steps between bunches. While there is an abrupt change in the terrace length at the upper edge of the bunch, making the transition with the low step-density region clear cut, at the lower edge, steps gradually depart from the bunch, blurring this transition zone, as observed in Fig.~\ref{fig:height_profile}. This asymmetrical distribution of steps around a bunch is not specific to our thermodynamically consistent model. It is also observed in the case of ES-triggered step bunching under evaporation \cite{Krug2005_scaling}, in the generic model \cite{Slanina2005} where the step velocity is a linear combination of the neighboring terrace widths, and in the Cellular Automaton-based model \cite{Krzyzewski2017}. 

We verify however that the bunch height is only weakly impacted: as all steps have the same height, the total bunch height is not dramatically modified by a few additional steps at the boundaries of the bunch, especially for large bunches. On the other hand, since terraces further from the bunch center are much wider than terraces close to it, the same additional steps have a considerable impact on the total bunch width. Hence, $W$ is strongly conditioned by the choice of threshold and cannot serve as a reliable indicator of the coarsening process. Lastly, as the distance between two bunches can be precisely determined as the distance between their respective sharp upper edges, $N$ is also a robust quantity. 

\begin{figure}[t]
\centering
	\includegraphics[width=0.4\textwidth]{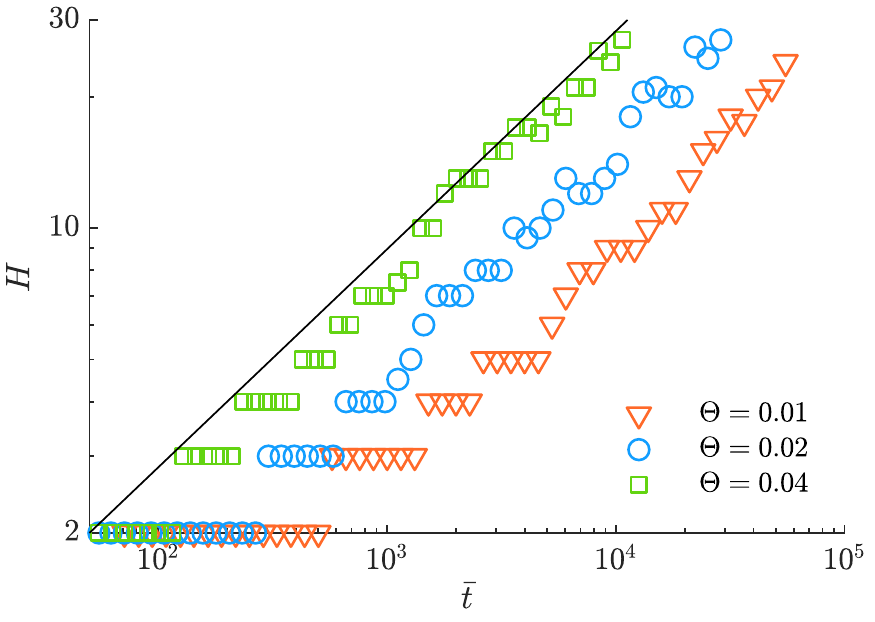}
	\vspace{\myPadding}
	\caption{Bunch height $H$ scaling with time $\ol{t}$. The black trend line shows the theoretical $H \sim t^{1/2}$ scaling. Same parameters as Fig. \ref{fig:spatiotemporal_diagram}.}
	\label{fig:scaling_H}
\end{figure}

\begin{figure}[t]
\centering
	\includegraphics[width=0.4\textwidth]{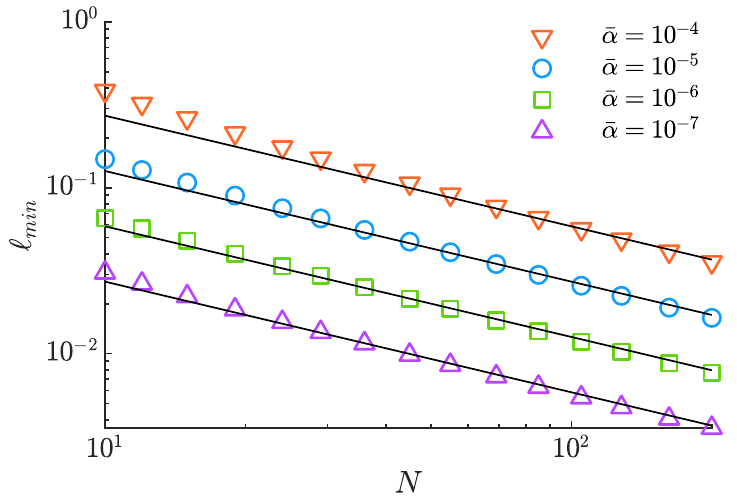}
	\vspace{\myPadding}
	\caption{Minimal interstep distance $\ell_{min}$ scaling with $N$. The black trend lines show the theoretical predictions ${\ell_{min}\sim N^{-2/3}}$. Same parameters as Fig. \ref{fig:spatiotemporal_diagram}.}
	\label{fig:scaling_lmin}
\end{figure}

The typical evolution of $H(\ol{t})$ and $\ell_{min}(N)$ are plotted in Fig.~\ref{fig:scaling_H} and \ref{fig:scaling_lmin}, where steps are considered as bunched when their distance is smaller than the initial terrace width \cite{Tonchev2012}. It is interesting to comment on the nondimensional physical parameters of our model. Based on experiments \cite{Chung2002,Ichimiya2000}, the kinetics of deposition on Si(111) at low temperatures (less than $900^{\degree}$C) is expected to be KL ($\ka \ll 1$) for miscut angles greater than $0.2^{\degree}$, and we thus restrict the parameter space to $\ka \leq 10^{-1}$. To obey the near-equilibrium hypothesis, additional restrictions are necessary. From the steady-state solution of \eqref{eq:ale}, the maximum departure of the adatom density from its equilibrium value can be estimated as $\max(\Fb/8,\Fb/\ka)$ so that the near-equilibrium hypothesis imposes $\Fb \ll 10$ and $\Fb \ll \ka$. Hence, the latter condition being more restrictive here, $\Fb \leq \ka/10$ is assumed in our simulations.

In addition, due to the low temperatures, we assume a low equilibrium adatom coverage $\Theta\!=\!0.02$. This value is conservatively low in the sense that it minimizes the strength of the dynamical and chemical effects. Indeed, estimates from the literature place this value closer to $0.04$ for Si \cite{Yang1994} and as high as $0.2$ for GaAs \cite{Johnson1996}. Finally, the elastic coefficient is set to $\ael\!=\! 10^{-5}$. With these restrictions, systematic simulations were conducted every decade for $\Fb$ from $10^{-5}$ to $10^{-2}$ and $\ka$ from $10^{-4}$ to $10^{-1}$. Additional simulations were conducted with $\Fb\!=\!10^{-4}$ and $\ka\!=\!10^{-2}$ for various values of $\Theta$ and $\ael$ in order to ascertain the scaling with respect to those parameters. From all the simulations performed, robust scaling laws emerge, and we find, in the absence of any ES barrier ($S\!=\!1$):
\begin{equation}\label{eq:scalings}
H \simeq 2.2\hspace{0.1em}\Theta^{0.7 \pm 0.05}\,\ol{t}^{1/2},\hspace{0.5em} \ell_{min} \simeq 1.6\Big(\frac{\ka\hspace{1pt}\ael}{\Fb\Theta}\Big)^{\!\! 1/3}N^{-2/3}.
\end{equation}

Systematic quantitative experiments on coarsening without electromigration are scarce. Of the three studies found in the literature, one concerns Si(001) \cite{Schelling2000}, on which adatom diffusion is strongly anisotropic, and another concerns GaAs(001) \cite{Ishizaki1996}, where the surface is grown by metalorganic vapor phase epitaxy, so that precursor interactions need to be taken into account. Since we are interested in investigating the influence of the dynamical and chemical effects on the step bunching instability, effects that are basic to step flow in the sense that they are present irrespective of whether adatom diffusion is anisotropic or not and whether chemical reaction between distinct species occur or not, we only consider Si(111) \cite{Omihomma2005}, which is the ideal candidate to test our model due to its isotropy and weak (or absent) ES barrier. On this surface, the bunch height and width were monitored and found to grow as $t^{\beta}$ and $t^{1/\alpha}$ with $\beta\!=\!0.49\!\pm\!0.09$ and $1/\alpha\!=\!0.54\!\pm\!0.08$. In the framework of universality classes based on the classical BCF model \cite{PTVV2002}, the destabilizing mechanism leading to the closest match ($\beta\!=\!1/\alpha\!=\! 0.5$) is the iES effect, whose existence remains controversial \cite{Slanina2005,Pimpinelli2000,Vladimirova2001}, with contradictory experimental results \cite{Chung2002,Voigtlander1995,Ichimiya1996,Rogilo2013}. Importantly, our simulations of the thermodynamically consistent model reproduce the bunch height scaling ($\beta\!=\!1/2$) without recourse to an iES ($S\!=\!1$). 

Since there are no experimental studies for the scaling of $\ell_{min}$ in the absence of electromigration, we are not able to test our prediction. However, we note that the exponent we find is identical to the one obtained in the simulations of \cite{Popkov_Krug_moving_bunches}.

\section{Continuum evolution equation}\label{sec:continuum_limit}
\subsection{Discrete-to-continuum limit}
In this section, we derive the continuum limit of the discrete step-flow equations \eqref{eq:ale}, \eqref{BC1d}, \eqref{BC2d} and \eqref{Vd}, whereby the stepped surface profile is described by a smooth function. The nonlinear PDE that governs the evolution of the surface height provides insight into the  mechanisms responsible for step bunching and explains the coarsening behavior observed in the simulations.

Two methods have been proposed in the literature for this discrete-to-continuum transition. In \cite{Krug2005_scaling}, a hybrid approach is adopted in which the nonlinear repulsion term is treated using a first-order correspondence between finite differences and derivatives, and the remaining terms, which form a linear combination of the adjacent terrace widths in the model considered, are coarse-grained through a Fourier transform \cite{Krug_homogenization}. In \cite{Xiang2002,Margetis2005}, the discrete step velocity equation is interpreted as a numerical scheme for a differential equation, with the step height representing the grid constant, in a process reminiscent of the modified equation technique \cite{Warming1974}. 

The derivation we present here is conducted in the same spirit as the latter method, but incorporates the dynamical and chemical effects, which are unaccounted for in the cited works, and ensures all terms are expanded to the same order in their Taylor-series representation. The expansion is based on the assumption that the terrace widths are small compared to the mesoscopic length scale $\LL$ characterizing the spatial variations of step density on the vicinal surface \cite{Margetis2006}. Let $\eps\!=\!\LO/\LL$ be the nondimensional parameter for the Taylor expansion. All functions and their derivatives are assumed to be bounded, i.e., $\bigO(1)$. 

Since equations \eqref{eq:ale}, \eqref{BC1d}, \eqref{BC2d} and \eqref{Vd} constitute a free-boundary problem with time-dependent coefficients that cannot be solved analytically, we will only retain the \emph{advective} contributions to the dynamical effect by making the change of variables
\begin{equation}
\hat{x} := \ol{x} - \ol{x}_n(\ol{t}) \text{ and } \rnh(\ol{t},\hat{x}) := \rnt(\ol{t},\ol{x}),
\end{equation}
and neglecting the partial time derivative of $\rnh$, so that \eqref{eq:ale} reduces to
\begin{equation}\label{eq:diff_eq_approx}
0 = \partial^2_{\hat{x}\hat{x}}\rnh + \chi_a \Pc \dot{\ol{x}}_n \partial_{\hat{x}}\rnh + \Fb.
\end{equation}
Introducing $u\!:=\!\hat{x}/\ol{s}_n$, the solution of \eqref{eq:diff_eq_approx} can be expressed as
\begin{equation}\label{rho_sol}
\rnh(\ol{t},u) = \rnh^+ \varphi_n(\ol{t},u) + \rnh^- \psi_n(\ol{t},u) + c_n(\ol{t},u),
\end{equation}
where the expressions for $\varphi_n(\ol{t},u)$, $\psi_n(\ol{t},u)$ and $c_n(\ol{t},u)$ are given in Appendix \ref{app:rho_sol}. 

In the boundary conditions \eqref{BC1d} and \eqref{BC2d}, the chemical effect couples the diffusion fields on adjacent terraces. We use the interface motion equation \eqref{Vd} to express $\hat{\rho}_{n+1}^+$ as a function of $\dot{\ol{x}}_{n+1}$ and $\rnh^-$, and $\hat{\rho}_{n-1}^-$ as a function of $\dot{\ol{x}}_n$ and $\rnh^+$. Inserting these expressions in \eqref{BC1d} and \eqref{BC2d}, and appealing to \eqref{rho_sol} to express the derivatives of $\rnh$ in terms of $\rnh^+$ and $\rnh^-$, we obtain a linear system that can be solved for $\rnh^+$ and $\rnh^-$ (whose explicit expressions are given in Appendix \ref{app:rho_sol}).

Substituting the resulting expressions in \eqref{Vd}, we obtain:
\begin{align}\label{eq:step_velocity}
\begin{split}
\Pc\dot{\ol{x}}_n = &\Pc\Big[ \frac{\ol{s}_n+\ol{s}_{n-1}}{2} + \frac{C_2-C_1}{2}\delta\Big( \frac{\ol{s}_{n-1}}{B_{n-1}} \Big) \Big] \\
- &\ka S \Theta \, \delta\Big( \frac{\delta(\ol{\mathfrak{f}}_{n-1})}{B_{n-1}} \Big) + \Pc\Theta C_0 \delta\Big( \frac{\delta(\dot{\ol{x}}_{n-1})}{B_{n-1}} \Big)\\
- &\Pc\Theta (\chi_a-\chi_c)\frac{S+1}{2}\delta\Big( \frac{\dot{\ol{x}}_n+\dot{\ol{x}}_{n-1}}{B_{n-1}} \Big) \\
+ &\bigO(\Pc^2,\Pc\ael),
\end{split}
\end{align}
where, for any $z_n, \delta(z_n)\!=\!z_{n+1}-z_n$, and
\begin{empheq}[left=\empheqlbrace]{align}
\begin{aligned}
B_n &:= 1+S+\ka S \ol{s}_n, \\
C_0 &:= (1-S)(\chi_a+\chi_c)/2 + \chi_a\chi_c\Theta(S+1), \\
C_1 &:= 1 + \chi_c\Theta(S+1), \\
C_2 &:= S - \chi_c\Theta(S+1).
\end{aligned}
\end{empheq}

In the KL regime, $\ka \ll 1$, so that $B_n \simeq 1 + S$. We can now proceed to the homogenization of \eqref{eq:step_velocity}. To approximate the profile of the vicinal surface, we introduce the continuous function $X(t,y)$ such that
\begin{equation}
\ol{x}_n(\ol{t}) = \eps^{-1} X(t = \eps \ol{t}, y=-n\eps),
\end{equation}
where the space variable has been normalized and the time appropriately rescaled to reflect the change from microscopic length scale $\LO$ to macroscopic length scale $\LL$. 
Taylor-expanding the different terms in \eqref{eq:step_velocity} up to order 3 (the dominant order of the elastic repulsion term), we get:
{
\small
\begin{empheq}[left=\empheqlbrace]{align}
\begin{aligned}
&\frac{\ol{s}_n+\ol{s}_{n-1}}{2} = -X_{y} - \frac{\eps^2}{6}X_{y^3} + \bigO(\eps^4), \\
&\delta\Big(\! \frac{\ol{s}_{n-1}}{B_{n-1}} \!\Big) = \frac{\eps}{S+1} \Big(X_{y^2} + \frac{\eps^2}{12} X_{y^4}\Big) + \bigO(\eps^5), \\
&\delta\Big(\! \frac{\delta \ol{\mathfrak{f}}_{n-1}}{B_{n-1}} \!\Big) = -\gamma(R)\frac{\ael\eps^3}{S+1} \Big[\frac{1}{X_{y}^3}\Big]_{y^3} + \bigO(\eps^5), \\
&\delta\Big(\! \frac{\delta(\dot{\ol{x}}_{n-1})}{B_{n-1}} \!\Big) = \frac{\eps^2}{S+1} X_{ty^2} + \bigO(\eps^4), \\
&\delta\Big(\! \frac{\dot{\ol{x}}_n+\dot{\ol{x}}_{n-1}}{B_{n-1}} \!\Big) = -\frac{2\eps}{S+1} \Big(X_{ty} + \frac{\eps^2}{6} X_{ty^3}\Big) + \bigO(\eps^5).
\end{aligned}
\end{empheq}
}

In the limit $R\!\rightarrow\!\infty$, $\gamma(R)\!=\!\sum_{r\!=\!1}^R{r^{-2}} \!\rightarrow\! \pi^2/6 \!\simeq\! 1.64 $. If the infinite sum is instead truncated at 5 terms like in the numerical simulations, $\gamma(5) \simeq 1.46$. Although this introduces an error of $11\%$, it has effectively no impact on the scaling law for the bunch height as $H$ is independent of the strength of the elastic repulsion.

\begin{figure}[t]
\vspace{-2em}
\centering
\hspace*{-1cm}
	\includegraphics[trim=15 0 38 0,clip,width=0.45\textwidth]{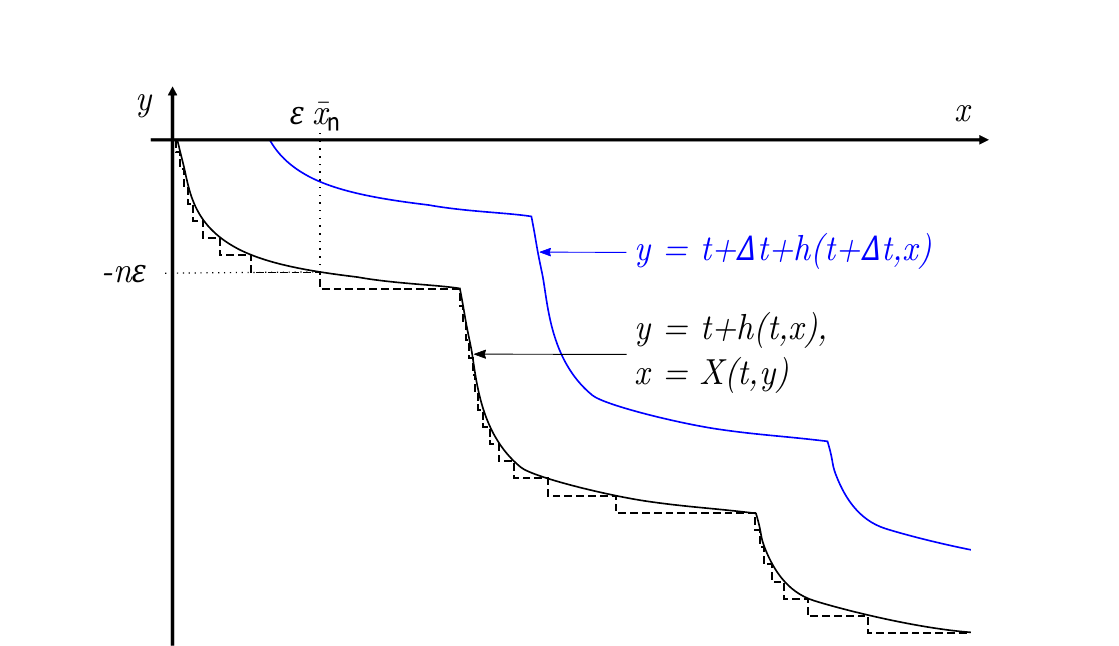}
	\vspace{\myPadding}
	\caption{In black, correspondence between the discrete and continuous surface profiles at time $t$. In blue, the surface profile at a later time $t+\Delta t$.}
	\label{fig:continuum_limit_profile}
\end{figure}

In order to obtain an equation for the the surface height $h(t,x)$, we start by introducing the nonlinear transform $t + h(t,\eps \ol{x}_n(\ol{t}))\!=\!-n\eps$, i.e., $t + h(t,X(t,y))\!=\!y$ (Fig.~\ref{fig:continuum_limit_profile}), such that we absorb the constant deposition term, yielding:
\begin{align}
X_t &= -\frac{1 + h_t}{h_x},\ X_{y} = \frac{1}{h_x},\ X_{yy} = \frac{1}{h_x}\Big[\frac{1}{h_x}\Big]_x,
\end{align}
and so on for higher-order derivatives. To eliminate the cross-derivative terms, we rearrange the terms of the equation by repeated differentiation, division by $h_x$, and substitutions, resulting in the sought-after PDE:
\begin{align}\label{eq:final_eq}
\begin{split}
h_t &- \eps K_1\Big[\frac{1}{h_x}\Big]_x + \eps^3 K_2 \Big[ \frac{1}{h_x}(h_x^2)_{xx} \Big]_x \\
&+ \eps^2 K_4\Big[ \frac{h_{xx}}{h_x^3} \Big]_x + \eps^3 K_5 \Big[ \frac{1}{h_x}\Big[ \frac{h_{xx}}{h_x^3} \Big]_x\Big]_x = \bigO(\eps^4),
\end{split}
\end{align}
where, having introduced ${K_0\!:=\!C_0/(S+1)}$ and ${K_3\!:=\!\dfrac{1}{12}\Big[(2\chi_a-\chi_c)\Theta-\dfrac{1}{2}\dfrac{S-1}{S+1} \Big]}$,
\begin{empheq}[left=\empheqlbrace]{align}
\begin{aligned}
K_1 &:= \chi_a\Theta - \frac{1}{2}\frac{S-1}{S+1}, \\
K_2 &:= \frac{3}{2}\frac{S}{S+1}\gamma(R)\frac{\ka\hspace{1pt}\ael}{\Fb}, \\
K_4 &:= \frac{1}{6}+\Theta K_0 + K_1\Theta(\chi_a-\chi_c), \\
K_5 &:= K_3 + \Theta K_0 K_1 + K_4\Theta(\chi_a-\chi_c).
\end{aligned}
\end{empheq}

The prevailing equation in the literature \cite{PTVV2002,Krug2005}, based on the quasistatic BCF model, can be recovered from Eq.~\eqref{eq:final_eq} by neglecting the dynamical and chemical effects ($\chi_a\!=\!\chi_c\!=\!0$) and setting $K_5\!=\!0$. While the first condition ensues naturally from the definitions of $\chi_a$ and $\chi_c$, the second amounts to neglecting a term that is of the same order as the $K_2$ term of elastic repulsion, which is not justified \emph{a priori}. Numerical integration of Eq.~\eqref{eq:final_eq} in the presence and absence of the $K_5$ term show that its impact on the bunch profile is limited to narrow regions at the upper and lower edges, with no visible effect on the bunch shape. Nevertheless, we show in the next section that the $K_5$ term plays an important role in the onset of instability, where its influence cannot be neglected.

\subsection{Linear stability analysis}
The linear-stability analysis of \eqref{eq:final_eq} is performed by setting $h(t,x)\!=\!-x + \delta h \, e^{ikx+\omega t}$, where $h(t,x)\!=\!-x$ corresponds to the fundamental solution, and expanding \eqref{eq:final_eq} to linear order in $\delta h$. Time and space are then rescaled ($\ol{k}\!=\!\eps k$ and $\ol{\omega}\!=\!\eps\omega$) for comparison with the discrete system, yielding the dispersion relation
\begin{equation}\label{dispersion}
\Re(\ol{\omega}) = K_1 \ol{k}^2 - (2K_2+K_5)\ol{k}^4.
\end{equation}

We conclude from \eqref{dispersion} that a step-bunching instability exists as long as ${K_1\!>\!0}$. For the quasistatic BCF model, $K_1\!=\!(S-1)/2(S+1)$, so that an iES barrier ($S\!<\!1$) is necessary to fulfill that condition, and its absence (${S\!=\!1}$) or the presence of a direct ES barrier (${S\!>\!1}$) leads to a stable step flow \cite{Schwoebel1966}. In contrast, the inclusion of the dynamical and chemical effects renders the recourse to an iES barrier unnecessary to explain instability, as long as the attachment-detachment asymmetry satisfies ${S\!<\!(1+2\Theta)/(1-2\Theta)}$. 

Note that setting ${S\!=\!(1-2\Theta)/(1+2\Theta)}$ in the quasistatic BCF model mimics the same $K_1$ coefficient than setting $S\!=\!1$ in the full model with the dynamical effect. In other words, the dynamical effect may be interpreted as an \emph{effective} iES effect, analogously to chemical reactions \cite{Pimpinelli2000} or diffusion anisotropy \cite{Schelling2000}. This interpretation also sheds a new light on the experimental uncertainty surrounding the nature of the ES barrier on Si(111). Indeed, the smallness of $\Theta$ implies a weak effective iES barrier, and since only indirect methods are available to determine this value, it is likely that the measurement accuracy is insufficient to conclude.

Regarding the chemical effect, the complete linear-stability analysis \cite{Guin_thesis} shows that its destabilizing effect is strongest for the step pairing mode but that its impact is reduced in the limit of large wavelengths, which is the relevant one when passing to the continuum limit, thus explaining its absence from the dominant destabilizing contribution in \eqref{dispersion}.

Going back to the discrete equation \eqref{eq:step_velocity} and setting $\ol{x}_n\!=\!n + \ol{t} + \delta x \, e^{i \ol{k} n+\ol{\omega}\ol{t}}$, we obtain:
\begin{align}
\begin{split}
\ol{\omega} = i \sin(\ol{k}) + \bigg( &4[K_1-(\chi_a-\chi_c)\Theta - K_0\Theta \, \ol{\omega}] \\
&-\Theta(\chi_a-\chi_c)i\sin(\ol{k})\,\ol{\omega} \\
&+32 \frac{\gamma_d(R,\ol{k})}{\gamma(R)}K_2 \bigg) \sin^2(\ol{k}/2),
\end{split}
\end{align}
with $\displaystyle \gamma_d(R,\ol{k}) := \sum\limits_{r=1}^R\frac{\sin^2(\ol{k}r/2)}{r^4}$.

Solving for $\ol{\omega}$, taking the real part, and expanding for long wavelengths ($\ol{k} \rightarrow 0$) up to $\bigO(\ol{k}^5)$, we recover the exact same expression \eqref{dispersion} as in the continuum limit. This confirms the validity and relevance of the continuum limit \eqref{eq:final_eq}, notably regarding the new $K_5$ term. Indeed, as $K_5 \gtrsim K_2$ for typical values of the model parameters, it has a significant influence on the maximum growth rate $\ol{\omega}_m\!=\!(2K_2+K_5)^{-1}(K_1/2)^2$ and the most unstable mode $\ol{k}_m\!=\!\sqrt{(2K_2+K_5)^{-1}K_1/2\,}$.

\section{Scaling laws}\label{sec:scaling_laws_theory}
There are two scaling laws of interest to describe the asymptotic behavior of the surface profile. The scaling of $H$ with time is an indicator of the evolution of surface roughness and the scaling of $\ell_{min}$ with $N$ characterizes the bunch shape. We also look at the bunch velocity $v$ scaling with $N$ as an additional descriptor of the coarsening process.
\subsection{Bunch height $H$}
A common approach to obtain scaling laws from PDE's relies on identifying self-similar solutions. In \cite{PTVV2002}, such solutions are introduced based on a simplified version of \eqref{eq:final_eq} where only the transient term $h_t$, the destabilizing $K_1$ term and the stabilizing $K_2$ term are considered. 

\begin{figure}[t]
\centering
	\includegraphics[width=0.4\textwidth]{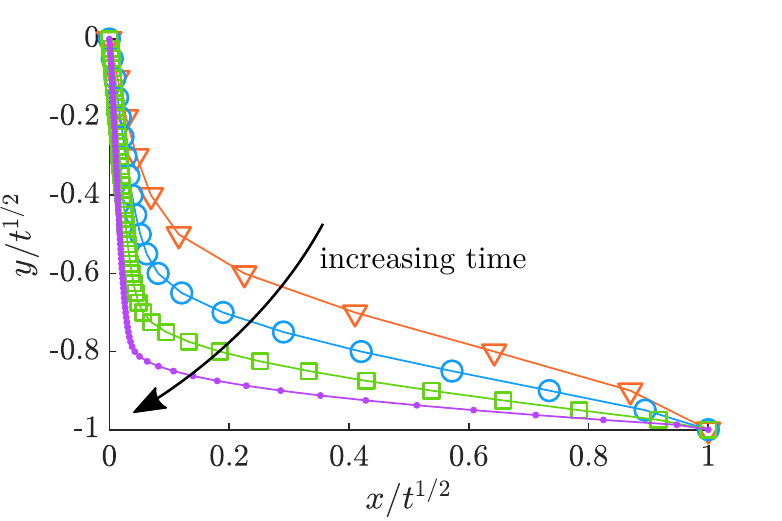}
	\vspace{\myPadding}
	\caption{Bunch steepening. Each curve corresponds to the rescaled bunch profile after a 4-fold time increase.}
	\label{fig:steepening}
\end{figure}

However, as pointed out in \cite{Krug2005_scaling,Slanina2005,Tonchev2012}, the obtained solutions do not reproduce the observed scaling laws. In each of these works, a different argument is invoked to justify the shortcomings of \cite{PTVV2002}. Here, we show that the reason is more fundamental: the evolution of the surface profile is simply not self-similar. Indeed, if we look for self-similar solutions of \eqref{eq:final_eq} of the form $h(t,x)\!=\!t^{a_1}\phi(\zeta)$, with $\zeta\!=\!x/t^{a_2}$, we find
\begin{align}\label{eq:self-similar}
\begin{split}
&t^{a_1-1}(a_1 \phi- a_2 \xi \phi') - \eps K_1 t^{-a_1} \Big[\frac{1}{\phi'}\Big]'\! \\
&\quad+ t^{-2 a_1}\eps^2 K_4\Big[\frac{\phi''}{\phi'^3}\Big]'\! + t^{a_1 - 4 a_2}\eps^3 K_2 \Big[ \frac{1}{\phi'}(\phi'^2)'' \Big]' \\
&\quad\quad+ t^{-3 a_1}\eps^3 K_5 \Big[ \frac{1}{\phi'}\Big[ \frac{\phi''}{\phi'^3}\Big]'\Big]'\! = 0.
\end{split}
\end{align}
Since this equation cannot be made scale invariant, it does not admit self-similar solutions. This is consistent with the profiles obtained from numerical simulations of the discrete step-flow equations which shows the steepening of the bunch despite rescaling it as per the expected $t^{1/2}$ scaling law (Fig.~\ref{fig:steepening}).

Analyzing \eqref{eq:self-similar} further, we note that the $K_4$ and $K_5$ terms, which preclude scale-invariance, present a $t^{-2a_1}$ and a $t^{-3a_1}$ factors, respectively. This indicates that they possibly become negligible at long times compared to the $K_1$ term exhibiting a $t^{-a_1}$ factor, provided that the associated functions $(\phi''/\phi'^3)'$ and $(1/\phi'(\phi''/\phi'^3)')'$ are regular enough. While this is the case inside the bunched and quasiflat regions, at the transition zones (which become sharper as the surface coarsens), these functions diverge, and the associated terms cannot be neglected, precluding the existence of self-similar solutions. 

Nonetheless, it seems clear from Fig.~\ref{fig:steepening} that the $t^{1/2}$ scaling  plays a crucial role, regardless of self-similarity. To see this, we modify the existing analysis to focus on the quasiflat region, so that the $K_4$ and $K_5$ terms may be neglected, and we consider asymptotic expansions for the characteristic height and length of the region of the form:
\begin{align}
\begin{split}
h(t,x) &= h_0(t,x) + \sum_{i \geq 0} t^{a_i}\varphi_i(\zeta), \\
\zeta &= \frac{x}{\sum_{j \geq 0} c_j t^{b_j}},
\end{split}
\end{align}
where $h_0(t,x)\!=\!-x$ represents the fundamental solution (equidistant steps), and $\forall i \in \mathbb{N}$, $a_i\!>\!a_{i-1}$, $b_i\!>\!b_{i-1}$, $\varphi_i(\zeta)\!=\!\bigO(1)$ and $c_0\!=\!1$. Thus,
\begin{equation}
h_x = -1 + \frac{\sum_{i \geq 0} t^{a_i}\varphi_i'(\zeta)}{\sum_{j \geq 0} c_j t^{b_j}}.
\end{equation}

Since the slope in the quasiflat region must remain finite as ${t\rightarrow\infty}$, $a_0 \leq b_0$ must hold. Moreover, as it cannot coincide with the $-1$ slope of the stable solution, strict inequality is not possible, so that $a_0\!=\!b_0$. Next, steps being far apart in the quasiflat region, we neglect the elastic term, and \eqref{eq:final_eq} becomes:
\begin{align}
\begin{split}
\sum_{i \geq 0}t^{a_i-1}&\Bigg[ a_i \varphi_i - \frac{\sum_{j \geq 0} b_j c_j t^{b_j}}{\sum_{j \geq 0} c_j t^{b_j}}\zeta\varphi' \Bigg] \\
&- \frac{\eps K_1}{\sum_{j \geq 0} c_j t^{b_j}}\Bigg[ \frac{1}{-1 + \frac{\sum_{i \geq 0} t^{a_i}\varphi_i'(\zeta)}{\sum_{j \geq 0} c_j t^{b_j}}} \Bigg]' = 0.
\end{split}
\end{align}
Looking at the dominant contribution, we get:
\begin{equation}
a_0 t^{a_0-1}\Big[\varphi_0 - \zeta \varphi'_0\Big] - \eps K_1 t^{-a_0}\Big[\frac{1}{-1+\varphi'_0}\Big]' = 0.
\end{equation}
Hence, scale invariance imposes $a_0-1\!=\!-a_0\!=\!-1/2$, and we recover the scaling law reported in the literature \cite{Krug2005_scaling,Omihomma2005}. 

\subsection{Minimal terrace size $\ell_{min}$}
As the previous analysis is conducted in the quasiflat region, a different approach is needed to determine the scaling law for $\ell_{min}$ in the bunch. In the stationary regime, the scaling of the bunch can be well approximated \cite{Krug2005_scaling,Stoyanov2000} and leads, for large enough bunches, to
\begin{equation}\label{eq:scaling_lmin}
\ell_{min} \simeq \Big(\frac{16}{3}\frac{K_2}{K_1}\Big)^{1/3}\! N^{-2/3} = 4^{1/3}\Big(\frac{\ka\hspace{1pt}\ael}{\Fb\Theta}\Big)^{1/3}\! N^{-2/3}.
\end{equation}
This expression predicts exactly the different exponents observed for each physical parameter and the theoretical prefactor $4^{1/3} \simeq 1.58$ is in excellent agreement with the $1.6$ numerical estimate found in \eqref{eq:scalings}.

However, although the scaling behavior of $\ell_{min}$ is accurately described in the context of the stationary approximation, a closer inspection reveals that the predicted slope of the bunch is symmetric with respect to its center, in disagreement with previous simulations \cite{Popkov_Krug_moving_bunches} and our own. Specifically, even though the velocity of a bunch decreases with its size, which \emph{a priori} legitimizes the stationary approximation for large bunches, it still has a crucial influence on the bunch shape. Indeed, if the bunch velocity is included, while neglecting other dynamical contributions, the expected asymmetric bunch shape is recovered \cite{Popkov_Krug_moving_bunches}. Nonetheless, as this adjustment only introduces a $6\%$ correction \cite{Popkov_Krug_moving_bunches} in the numerical prefactor of \eqref{eq:scaling_lmin}, its validity can thus be extended from the stationary to the quasisteady regime.

\subsection{Bunch velocity}

In this section, we analytically derive an expression for the bunch velocity, which was previously only assessed via numerical simulations \cite{Popkov_Krug_moving_bunches}. Neglecting the $K_5$ term in equation \eqref{eq:final_eq}, we apply the traveling-wave change of variable $h(t,x)\!=\!g(x-vt) + \Omega t$ to transform the PDE into an ODE. Denoting derivatives with respect to $\xi\!=\!x-vt$ by primes, we obtain:
{
\small
\begin{equation}\label{eq:traveling-wave}
-\!v(1+g') - \eps K_1 \Big[\frac{1}{g'}\Big]'\!\! + \eps^3 K_2 \Big[\frac{(g'^2)''}{g'}\Big]'\!\! + \eps^2 K_4\Big[\frac{g''}{g'^3}\Big]'\!\! = 0,
\end{equation} 
}
after identifying $\Omega\!=\!-v$ from the fundamental solution for which $g'\!=\!-1$.

\begin{figure}[b]
\centering
	\includegraphics[width=0.45\textwidth]{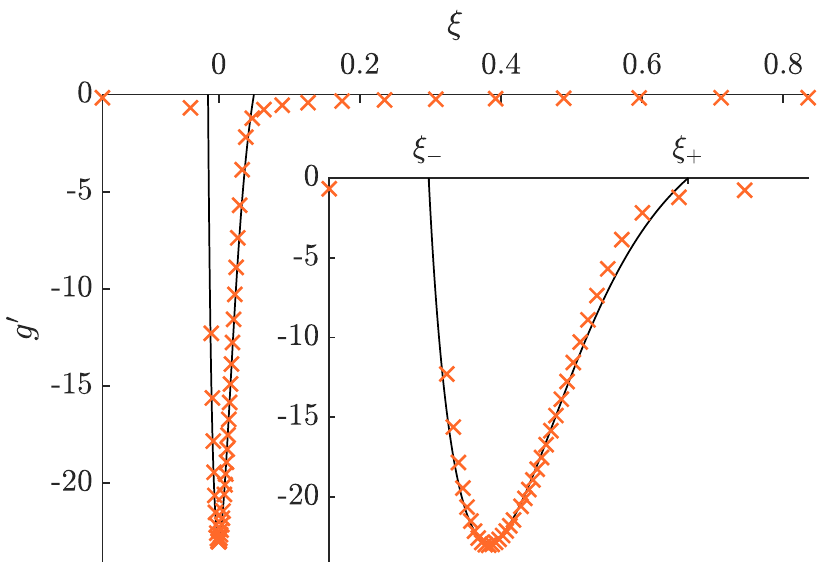}
	\vspace{\myPadding}
	\caption{Comparison of the surface slope obtained from simulations of the discrete system under forced bunching (x) with the analytic expression \eqref{eq:pade_approximant}, where the coefficients are given by \eqref{eq:pade_coeff} and $v$ is determined from the numerical resolution of \eqref{eq:traveling-wave}. The inset focuses on the bunched region.}
	\label{fig:pade_profile}
\end{figure}

Denoting $M$ the absolute value of the maximum slope in the bunch, we use a (2,3) Padé approximant to estimate the shape of the bunch slope:
\begin{equation}\label{eq:pade_approximant}
g'(\xi) = \frac{-M + a_1 \xi + a_2 \xi^2}{1 + b_1\xi + b_2\xi^2 + b_3\xi^3},
\end{equation}
where the $\xi$-origin is set at the point of maximum slope. The condition $g''(0)\!=\!0$ imposes $b_1\!=\!-a_1/M$. The remaining coefficients $a_1, a_2, b_2$, and $b_3$ are determined following the procedure detailed in Appendix \ref{app:pade_coeff}. Setting $\eta\!:=\!v/(\eps^3 K_2)$, we obtain, at leading order:
\begin{empheq}[left=\empheqlbrace]{align}\label{eq:pade_coeff}
\begin{aligned}
&& a_1 &\simeq A_1\eta^{1/3}M,\ && a_2 \simeq A_2\eta^{2/3}M, \\ 
&& b_2 &\simeq B_2\eta^{2/3},\ && b_3 \simeq B_3\eta,
\end{aligned}
\end{empheq}
The exact expressions for the $A_i$ and $B_i$ are not reported, as they consist of tedious polynomial roots with no special interest.

To compute the relationship between $M$ and $v$, we use the fact that the height of the bunch is normalized to~1. When integrating the slope, we neglect the contribution of the quasiflat terraces and assume that the main contribution comes from the bunched region, i.e., the region between the roots $\xi_-$ and $\xi_+$ of $g'$ (Fig.~\ref{fig:pade_profile}). Hence:
\begin{equation}
\int_{\xi_-}^{\xi_+} g'(\xi) \ud \xi \simeq -1.
\end{equation}

As the exact integration is unnecessarily laborious, we use a third-order Gauss quadrature (higher orders procure negligible corrections) to get an approximate expression. The dominant contribution yields:
\begin{equation}
 M \simeq 0.303 \, \eta^{1/3}.
\end{equation}

\begin{figure}[t]
\centering
	\includegraphics[width=0.4\textwidth]{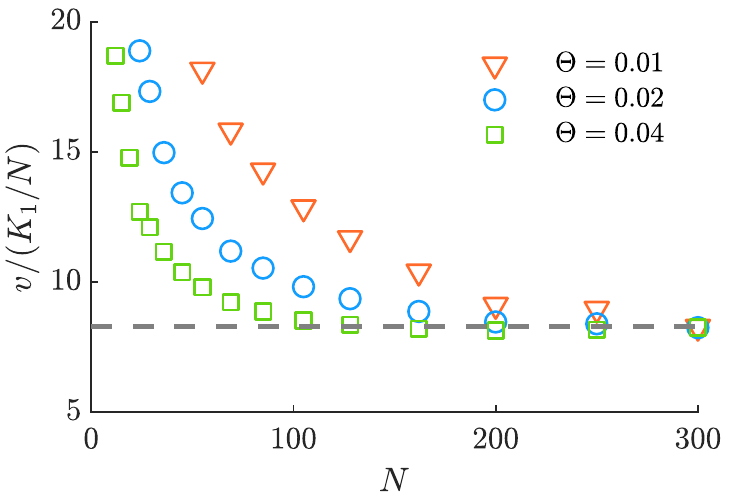}
	\vspace{\myPadding}
	\caption{Bunch velocity $v$ scaling with $N$. The asymptote is at 8.3.}
	\label{fig:scaling_v}
\end{figure}

Recalling that the velocity of a moving bunch mainly impacts its shape but has a negligible effect on the maximum slope $M$ \cite{Popkov_Krug_moving_bunches}, we finally conclude:
\begin{equation}
0.303^3 \frac{v}{\eps^3 K_2} \simeq \frac{3}{16} \frac{K_1}{\eps^2 K_2},
\end{equation}
yielding the expression
\begin{equation}\label{eq:analytic_v}
v \simeq 6.74 \frac{K_1}{N}.
\end{equation}
While we correctly predict the scaling $v \sim K_1/N$, in agreement with \cite{Popkov_Krug_moving_bunches} and our own simulations, the prefactor is 20\% smaller than the expected value of 8.3 derived from the numerical simulations (Fig.~\ref{fig:scaling_v}). This can be traced back to the fact that the Padé approximant does not capture the exact bunch shape (Fig. \ref{fig:continuum_limit_profile}). In addition, the error is also expanded by the cubic power applied in \eqref{eq:analytic_v}.

\section{Summary and discussion}
In this paper, we revisit the coarsening behavior of step bunching on vicinal surfaces in the kinetic-limited growth regime, with our starting point a thermodynamically consistent generalization of the BCF model that accounts for the dynamics of adatom diffusion on terraces and adatom attachment-detachment at steps, and the necessary coupling of the diffusion fields on adjacent terraces. Our numerical simulations show that these dynamical and chemical effects can account for the onset of step bunching and for the scaling laws observed in the coarsening regime, thereby circumventing the uncertainty surrounding the existence of an inverse Ehrlich-Schwoebel barrier.

Through a careful rescaling and systematic Taylor expansions, we propose a coherent discrete-to-continuum derivation, leading to a nonlinear PDE that describes the macroscopic evolution of the surface profile. This continuum limit differs from those found in the literature in that it incorporates the dynamical and chemical effects. The contributions of these effects to the coefficients of the nonlinear equation show that the step-bunching instability may be triggered even in the presence of a direct Ehrlich-Schwoebel barrier, in contrast to the conclusions drawn in the framework of the quasistatic approximation where an inverse Ehrlich-Schwoebel barrier is required. Moreover, we report a new term, the $K_5$ term in \eqref{eq:final_eq}, which, given its magnitude, has a crucial influence on the growth rate and the most unstable mode of the instability.

We also show, using the derived continuum limit, that the evolution of bunches is not self-similar, in agreement with the numerical simulations of the discrete step-flow equations, in contrast with the results found in the literature, based on a simplified, and thus incomplete, evolution equation. However, by taking into account the multiple length scales at play and by restricting our scaling analysis to the quasiflat region of the bunch cell to avoid the high-curvature transition zones with diverging terms, our investigation recovers the $1/2$ coarsening exponent for the bunch height $H$, which has been observed experimentally and reported in a number of simulations.

To derive the appropriate stationary scaling for the minimal interstep distance $\ell_{min}$, we transpose the quasisteady analysis of \cite{Popkov_Krug_moving_bunches} to our thermodynamically consistent model. Finally, we derive a theoretical expression for the bunch velocity, which was previously identified numerically, offering a way of determining the destabilizing factor $K_1$ from macroscopic features of the nonlinear evolution of the vicinal surface.

With this in mind, we conduct additional simulations of the discrete step-flow equations with various values of $S$ (not reported here) to determine the scaling of the bunch height and minimal interstep distance with $S$. The resulting generalized scaling laws are: ${H\!\sim\!2.2 K_1^{0.07} \bar{t}^{1/2}}$ and ${\ell_{min}\!\sim\!1.7(K_2/K_1)^{1/3}N^{-2/3}}$. Therefore, the identification of $K_1$ and $K_2$ from the prefactors of these scaling laws offers an interesting alternative for determining microscopic parameters of the vicinal surface (e.g., $\ka, \ael,$ and $\Theta$) from macroscopic features.

Our extension of the stability analysis of the thermodynamically consistent step-flow model to account for adatom electromigration (with special attention to the extreme-deposition regime, where the dynamical effect plays a crucial role in destabilizing the vicinal surface), and the derivation and analysis of the corresponding continuum limit will be presented elsewhere. Finally, we have generalized the present one-dimensional step-flow model to two space dimensions. The resulting analysis of the onset of the step meandering and of the coexistence of the bunching anf meandering instabilities is underway.

\section*{Acknowledgments}
L. Benoit-\phantom{}-Maréchal acknowledges the support of the \emph{\'{E}cole {P}olytechnique} through the AMX program financed by the \emph{Ministère de l’Enseignement Supérieur et de la Recherche et de l’Innovation}. The authors also wish to thank Dr. L. Guin for helpful discussions.

\bibliographystyle{apsrev4-2}
\bibliography{Scaling_Laws_Step_Bunching}

\begin{thebibliography}{60}%
\makeatletter
\providecommand \@ifxundefined [1]{%
 \@ifx{#1\undefined}
}%
\providecommand \@ifnum [1]{%
 \ifnum #1\expandafter \@firstoftwo
 \else \expandafter \@secondoftwo
 \fi
}%
\providecommand \@ifx [1]{%
 \ifx #1\expandafter \@firstoftwo
 \else \expandafter \@secondoftwo
 \fi
}%
\providecommand \natexlab [1]{#1}%
\providecommand \enquote  [1]{``#1''}%
\providecommand \bibnamefont  [1]{#1}%
\providecommand \bibfnamefont [1]{#1}%
\providecommand \citenamefont [1]{#1}%
\providecommand \href@noop [0]{\@secondoftwo}%
\providecommand \href [0]{\begingroup \@sanitize@url \@href}%
\providecommand \@href[1]{\@@startlink{#1}\@@href}%
\providecommand \@@href[1]{\endgroup#1\@@endlink}%
\providecommand \@sanitize@url [0]{\catcode `\\12\catcode `\$12\catcode
  `\&12\catcode `\#12\catcode `\^12\catcode `\_12\catcode `\%12\relax}%
\providecommand \@@startlink[1]{}%
\providecommand \@@endlink[0]{}%
\providecommand \url  [0]{\begingroup\@sanitize@url \@url }%
\providecommand \@url [1]{\endgroup\@href {#1}{\urlprefix }}%
\providecommand \urlprefix  [0]{URL }%
\providecommand \Eprint [0]{\href }%
\providecommand \doibase [0]{https://doi.org/}%
\providecommand \selectlanguage [0]{\@gobble}%
\providecommand \bibinfo  [0]{\@secondoftwo}%
\providecommand \bibfield  [0]{\@secondoftwo}%
\providecommand \translation [1]{[#1]}%
\providecommand \BibitemOpen [0]{}%
\providecommand \bibitemStop [0]{}%
\providecommand \bibitemNoStop [0]{.\EOS\space}%
\providecommand \EOS [0]{\spacefactor3000\relax}%
\providecommand \BibitemShut  [1]{\csname bibitem#1\endcsname}%
\let\auto@bib@innerbib\@empty
\bibitem [{\citenamefont {Ronda}\ and\ \citenamefont
  {Berbezier}(2004)}]{Ronda2004}%
  \BibitemOpen
  \bibfield  {author} {\bibinfo {author} {\bibfnamefont {A.}~\bibnamefont
  {Ronda}}\ and\ \bibinfo {author} {\bibfnamefont {I.}~\bibnamefont
  {Berbezier}},\ }\href
  {https://doi.org/https://doi.org/10.1016/j.physe.2003.12.136} {\bibfield
  {journal} {\bibinfo  {journal} {Physica E: Low-dimensional Systems and
  Nanostructures}\ }\textbf {\bibinfo {volume} {23}},\ \bibinfo {pages} {370}
  (\bibinfo {year} {2004})},\ \bibinfo {note} {proceedings of the Fifth
  International Workshop on Epitaxial Semiconductors on Patterned Substrates
  and Novel Index Surfaces (ESPS-NIS)}\BibitemShut {NoStop}%
\bibitem [{\citenamefont {Wise}\ \emph {et~al.}(2005)\citenamefont {Wise},
  \citenamefont {Lowengrub}, \citenamefont {Kim}, \citenamefont {Thornton},
  \citenamefont {Voorhees},\ and\ \citenamefont {Johnson}}]{Wise2005}%
  \BibitemOpen
  \bibfield  {author} {\bibinfo {author} {\bibfnamefont {S.~M.}\ \bibnamefont
  {Wise}}, \bibinfo {author} {\bibfnamefont {J.~S.}\ \bibnamefont {Lowengrub}},
  \bibinfo {author} {\bibfnamefont {J.~S.}\ \bibnamefont {Kim}}, \bibinfo
  {author} {\bibfnamefont {K.}~\bibnamefont {Thornton}}, \bibinfo {author}
  {\bibfnamefont {P.~W.}\ \bibnamefont {Voorhees}},\ and\ \bibinfo {author}
  {\bibfnamefont {W.~C.}\ \bibnamefont {Johnson}},\ }\href
  {https://doi.org/10.1063/1.2061852} {\bibfield  {journal} {\bibinfo
  {journal} {Applied Physics Letters}\ }\textbf {\bibinfo {volume} {87}},\
  \bibinfo {pages} {133102} (\bibinfo {year} {2005})}\BibitemShut {NoStop}%
\bibitem [{\citenamefont {Krug}(2005)}]{Krug2005}%
  \BibitemOpen
  \bibfield  {author} {\bibinfo {author} {\bibfnamefont {J.}~\bibnamefont
  {Krug}},\ }in\ \href {https://link.springer.com/book/10.1007/b137679} {\emph
  {\bibinfo {booktitle} {Multiscale Modeling in Epitaxial Growth}}},\ \bibinfo
  {editor} {edited by\ \bibinfo {editor} {\bibfnamefont {A.}~\bibnamefont
  {Voigt}}}\ (\bibinfo  {publisher} {Birkh{\"a}user Basel},\ \bibinfo {address}
  {Basel},\ \bibinfo {year} {2005})\ pp.\ \bibinfo {pages} {69--95}\BibitemShut
  {NoStop}%
\bibitem [{\citenamefont {Mysliveček}\ \emph {et~al.}(2002)\citenamefont
  {Mysliveček}, \citenamefont {Schelling}, \citenamefont {Schäffler},
  \citenamefont {Springholz}, \citenamefont {Šmilauer}, \citenamefont {Krug},\
  and\ \citenamefont {Voigtländer}}]{Myslivecek2002}%
  \BibitemOpen
  \bibfield  {author} {\bibinfo {author} {\bibfnamefont {J.}~\bibnamefont
  {Mysliveček}}, \bibinfo {author} {\bibfnamefont {C.}~\bibnamefont
  {Schelling}}, \bibinfo {author} {\bibfnamefont {F.}~\bibnamefont
  {Schäffler}}, \bibinfo {author} {\bibfnamefont {G.}~\bibnamefont
  {Springholz}}, \bibinfo {author} {\bibfnamefont {P.}~\bibnamefont
  {Šmilauer}}, \bibinfo {author} {\bibfnamefont {J.}~\bibnamefont {Krug}},\
  and\ \bibinfo {author} {\bibfnamefont {B.}~\bibnamefont {Voigtländer}},\
  }\href {https://doi.org/https://doi.org/10.1016/S0039-6028(02)02273-2}
  {\bibfield  {journal} {\bibinfo  {journal} {Surface Science}\ }\textbf
  {\bibinfo {volume} {520}},\ \bibinfo {pages} {193 } (\bibinfo {year}
  {2002})}\BibitemShut {NoStop}%
\bibitem [{\citenamefont {Latyshev}\ \emph {et~al.}(1989)\citenamefont
  {Latyshev}, \citenamefont {Aseev}, \citenamefont {Krasilnikov},\ and\
  \citenamefont {Stenin}}]{Latyshev1989}%
  \BibitemOpen
  \bibfield  {author} {\bibinfo {author} {\bibfnamefont {A.}~\bibnamefont
  {Latyshev}}, \bibinfo {author} {\bibfnamefont {A.}~\bibnamefont {Aseev}},
  \bibinfo {author} {\bibfnamefont {A.}~\bibnamefont {Krasilnikov}},\ and\
  \bibinfo {author} {\bibfnamefont {S.}~\bibnamefont {Stenin}},\ }\href
  {https://doi.org/https://doi.org/10.1016/0039-6028(89)90256-2} {\bibfield
  {journal} {\bibinfo  {journal} {Surface Science}\ }\textbf {\bibinfo {volume}
  {213}},\ \bibinfo {pages} {157} (\bibinfo {year} {1989})}\BibitemShut
  {NoStop}%
\bibitem [{\citenamefont {Stoyanov}(1991)}]{Stoyanov1991}%
  \BibitemOpen
  \bibfield  {author} {\bibinfo {author} {\bibfnamefont {S.}~\bibnamefont
  {Stoyanov}},\ }\href {https://doi.org/10.1143/jjap.30.1} {\bibfield
  {journal} {\bibinfo  {journal} {Japanese Journal of Applied Physics}\
  }\textbf {\bibinfo {volume} {30}},\ \bibinfo {pages} {1} (\bibinfo {year}
  {1991})}\BibitemShut {NoStop}%
\bibitem [{\citenamefont {Kandel}\ and\ \citenamefont
  {Weeks}(1994)}]{Kandel1994}%
  \BibitemOpen
  \bibfield  {author} {\bibinfo {author} {\bibfnamefont {D.}~\bibnamefont
  {Kandel}}\ and\ \bibinfo {author} {\bibfnamefont {J.~D.}\ \bibnamefont
  {Weeks}},\ }\href {https://doi.org/10.1103/PhysRevB.49.5554} {\bibfield
  {journal} {\bibinfo  {journal} {Phys. Rev. B}\ }\textbf {\bibinfo {volume}
  {49}},\ \bibinfo {pages} {5554} (\bibinfo {year} {1994})}\BibitemShut
  {NoStop}%
\bibitem [{\citenamefont {Pimpinelli}\ and\ \citenamefont
  {Videcoq}(2000)}]{Pimpinelli2000}%
  \BibitemOpen
  \bibfield  {author} {\bibinfo {author} {\bibfnamefont {A.}~\bibnamefont
  {Pimpinelli}}\ and\ \bibinfo {author} {\bibfnamefont {A.}~\bibnamefont
  {Videcoq}},\ }\href {https://doi.org/10.1016/s0039-6028(99)01100-0}
  {\bibfield  {journal} {\bibinfo  {journal} {Surface Science}\ }\textbf
  {\bibinfo {volume} {445}},\ \bibinfo {pages} {L23} (\bibinfo {year}
  {2000})}\BibitemShut {NoStop}%
\bibitem [{\citenamefont {Pierre-Louis}\ \emph {et~al.}(1999)\citenamefont
  {Pierre-Louis}, \citenamefont {D'Orsogna},\ and\ \citenamefont
  {Einstein}}]{Pierre-Louis1999}%
  \BibitemOpen
  \bibfield  {author} {\bibinfo {author} {\bibfnamefont {O.}~\bibnamefont
  {Pierre-Louis}}, \bibinfo {author} {\bibfnamefont {M.~R.}\ \bibnamefont
  {D'Orsogna}},\ and\ \bibinfo {author} {\bibfnamefont {T.~L.}\ \bibnamefont
  {Einstein}},\ }\href {https://doi.org/10.1103/PhysRevLett.82.3661} {\bibfield
   {journal} {\bibinfo  {journal} {Phys. Rev. Lett.}\ }\textbf {\bibinfo
  {volume} {82}},\ \bibinfo {pages} {3661} (\bibinfo {year}
  {1999})}\BibitemShut {NoStop}%
\bibitem [{\citenamefont {Murty}\ and\ \citenamefont
  {Cooper}(1999)}]{Murty1999}%
  \BibitemOpen
  \bibfield  {author} {\bibinfo {author} {\bibfnamefont {M.~V.~R.}\
  \bibnamefont {Murty}}\ and\ \bibinfo {author} {\bibfnamefont {B.~H.}\
  \bibnamefont {Cooper}},\ }\href {https://doi.org/10.1103/PhysRevLett.83.352}
  {\bibfield  {journal} {\bibinfo  {journal} {Phys. Rev. Lett.}\ }\textbf
  {\bibinfo {volume} {83}},\ \bibinfo {pages} {352} (\bibinfo {year}
  {1999})}\BibitemShut {NoStop}%
\bibitem [{\citenamefont {Guin}\ \emph {et~al.}(2020)\citenamefont {Guin},
  \citenamefont {Jabbour}, \citenamefont {Shaabani-Ardali}, \citenamefont
  {Benoit-Mar\'echal},\ and\ \citenamefont {Triantafyllidis}}]{Guin2020}%
  \BibitemOpen
  \bibfield  {author} {\bibinfo {author} {\bibfnamefont {L.}~\bibnamefont
  {Guin}}, \bibinfo {author} {\bibfnamefont {M.~E.}\ \bibnamefont {Jabbour}},
  \bibinfo {author} {\bibfnamefont {L.}~\bibnamefont {Shaabani-Ardali}},
  \bibinfo {author} {\bibfnamefont {L.}~\bibnamefont {Benoit-Mar\'echal}},\
  and\ \bibinfo {author} {\bibfnamefont {N.}~\bibnamefont {Triantafyllidis}},\
  }\href {https://doi.org/10.1103/PhysRevLett.124.036101} {\bibfield  {journal}
  {\bibinfo  {journal} {Phys. Rev. Lett.}\ }\textbf {\bibinfo {volume} {124}},\
  \bibinfo {pages} {036101} (\bibinfo {year} {2020})}\BibitemShut {NoStop}%
\bibitem [{\citenamefont {Cermelli}\ and\ \citenamefont
  {Jabbour}(2005)}]{Cermelli2005}%
  \BibitemOpen
  \bibfield  {author} {\bibinfo {author} {\bibfnamefont {P.}~\bibnamefont
  {Cermelli}}\ and\ \bibinfo {author} {\bibfnamefont {M.}~\bibnamefont
  {Jabbour}},\ }\href {https://doi.org/10.1098/rspa.2005.1495} {\bibfield
  {journal} {\bibinfo  {journal} {Proceedings of the Royal Society of London A:
  Mathematical, Physical and Engineering Sciences}\ }\textbf {\bibinfo {volume}
  {461}},\ \bibinfo {pages} {3483} (\bibinfo {year} {2005})}\BibitemShut
  {NoStop}%
\bibitem [{\citenamefont {Guin}\ \emph
  {et~al.}(2021{\natexlab{a}})\citenamefont {Guin}, \citenamefont {Jabbour},\
  and\ \citenamefont {Triantafyllidis}}]{Guin2021A}%
  \BibitemOpen
  \bibfield  {author} {\bibinfo {author} {\bibfnamefont {L.}~\bibnamefont
  {Guin}}, \bibinfo {author} {\bibfnamefont {M.~E.}\ \bibnamefont {Jabbour}},\
  and\ \bibinfo {author} {\bibfnamefont {N.}~\bibnamefont {Triantafyllidis}},\
  }\href@noop {} {\bibinfo {title} {Revisiting step instabilities on crystal
  surfaces. {P}art {I}: The quasistatic approximation}} (\bibinfo {year}
  {2021}{\natexlab{a}}),\ \Eprint {https://arxiv.org/abs/2101.02612}
  {arXiv:2101.02612 [cond-mat.mtrl-sci]} \BibitemShut {NoStop}%
\bibitem [{\citenamefont {Guin}\ \emph
  {et~al.}(2021{\natexlab{b}})\citenamefont {Guin}, \citenamefont {Jabbour},
  \citenamefont {Shaabani-Ardali},\ and\ \citenamefont
  {Triantafyllidis}}]{Guin2021B}%
  \BibitemOpen
  \bibfield  {author} {\bibinfo {author} {\bibfnamefont {L.}~\bibnamefont
  {Guin}}, \bibinfo {author} {\bibfnamefont {M.~E.}\ \bibnamefont {Jabbour}},
  \bibinfo {author} {\bibfnamefont {L.}~\bibnamefont {Shaabani-Ardali}},\ and\
  \bibinfo {author} {\bibfnamefont {N.}~\bibnamefont {Triantafyllidis}},\
  }\href@noop {} {\bibinfo {title} {Revisiting step instabilities on crystal
  surfaces. {P}art {II}: General theory}} (\bibinfo {year}
  {2021}{\natexlab{b}}),\ \Eprint {https://arxiv.org/abs/2101.02614}
  {arXiv:2101.02614 [cond-mat.mtrl-sci]} \BibitemShut {NoStop}%
\bibitem [{\citenamefont {Voigtlander}\ \emph {et~al.}(1995)\citenamefont
  {Voigtlander}, \citenamefont {Zinner}, \citenamefont {Weber},\ and\
  \citenamefont {Bonzel}}]{Voigtlander1995}%
  \BibitemOpen
  \bibfield  {author} {\bibinfo {author} {\bibfnamefont {B.}~\bibnamefont
  {Voigtlander}}, \bibinfo {author} {\bibfnamefont {A.}~\bibnamefont {Zinner}},
  \bibinfo {author} {\bibfnamefont {T.}~\bibnamefont {Weber}},\ and\ \bibinfo
  {author} {\bibfnamefont {H.~P.}\ \bibnamefont {Bonzel}},\ }\href
  {https://doi.org/10.1103/physrevb.51.7583} {\bibfield  {journal} {\bibinfo
  {journal} {Physical Review B}\ }\textbf {\bibinfo {volume} {51}},\ \bibinfo
  {pages} {7583} (\bibinfo {year} {1995})}\BibitemShut {NoStop}%
\bibitem [{\citenamefont {Ichimiya}\ \emph {et~al.}(1996)\citenamefont
  {Ichimiya}, \citenamefont {Tanaka},\ and\ \citenamefont
  {Ishiyama}}]{Ichimiya1996}%
  \BibitemOpen
  \bibfield  {author} {\bibinfo {author} {\bibfnamefont {A.}~\bibnamefont
  {Ichimiya}}, \bibinfo {author} {\bibfnamefont {Y.}~\bibnamefont {Tanaka}},\
  and\ \bibinfo {author} {\bibfnamefont {K.}~\bibnamefont {Ishiyama}},\ }\href
  {https://doi.org/10.1103/physrevlett.76.4721} {\bibfield  {journal} {\bibinfo
   {journal} {Physical Review Letters}\ }\textbf {\bibinfo {volume} {76}},\
  \bibinfo {pages} {4721} (\bibinfo {year} {1996})}\BibitemShut {NoStop}%
\bibitem [{\citenamefont {Chung}\ and\ \citenamefont
  {Altman}(2002)}]{Chung2002}%
  \BibitemOpen
  \bibfield  {author} {\bibinfo {author} {\bibfnamefont {W.~F.}\ \bibnamefont
  {Chung}}\ and\ \bibinfo {author} {\bibfnamefont {M.~S.}\ \bibnamefont
  {Altman}},\ }\href {https://doi.org/10.1103/PhysRevB.66.075338} {\bibfield
  {journal} {\bibinfo  {journal} {Phys. Rev. B}\ }\textbf {\bibinfo {volume}
  {66}},\ \bibinfo {pages} {075338} (\bibinfo {year} {2002})}\BibitemShut
  {NoStop}%
\bibitem [{\citenamefont {Rogilo}\ \emph {et~al.}(2013)\citenamefont {Rogilo},
  \citenamefont {Fedina}, \citenamefont {Kosolobov}, \citenamefont
  {Ranguelov},\ and\ \citenamefont {Latyshev}}]{Rogilo2013}%
  \BibitemOpen
  \bibfield  {author} {\bibinfo {author} {\bibfnamefont {D.~I.}\ \bibnamefont
  {Rogilo}}, \bibinfo {author} {\bibfnamefont {L.~I.}\ \bibnamefont {Fedina}},
  \bibinfo {author} {\bibfnamefont {S.~S.}\ \bibnamefont {Kosolobov}}, \bibinfo
  {author} {\bibfnamefont {B.~S.}\ \bibnamefont {Ranguelov}},\ and\ \bibinfo
  {author} {\bibfnamefont {A.~V.}\ \bibnamefont {Latyshev}},\ }\href
  {https://doi.org/10.1103/PhysRevLett.111.036105} {\bibfield  {journal}
  {\bibinfo  {journal} {Phys. Rev. Lett.}\ }\textbf {\bibinfo {volume} {111}},\
  \bibinfo {pages} {036105} (\bibinfo {year} {2013})}\BibitemShut {NoStop}%
\bibitem [{\citenamefont {\ifmmode~\check{S}\else \v{S}\fi{}milauer}\ and\
  \citenamefont {Vvedensky}(1995)}]{Smilauer1995}%
  \BibitemOpen
  \bibfield  {author} {\bibinfo {author} {\bibfnamefont {P.}~\bibnamefont
  {\ifmmode~\check{S}\else \v{S}\fi{}milauer}}\ and\ \bibinfo {author}
  {\bibfnamefont {D.~D.}\ \bibnamefont {Vvedensky}},\ }\href
  {https://doi.org/10.1103/PhysRevB.52.14263} {\bibfield  {journal} {\bibinfo
  {journal} {Phys. Rev. B}\ }\textbf {\bibinfo {volume} {52}},\ \bibinfo
  {pages} {14263} (\bibinfo {year} {1995})}\BibitemShut {NoStop}%
\bibitem [{\citenamefont {Krug}(1997{\natexlab{a}})}]{Krug1997}%
  \BibitemOpen
  \bibfield  {author} {\bibinfo {author} {\bibfnamefont {J.}~\bibnamefont
  {Krug}},\ }\href {https://doi.org/10.1080/00018739700101498} {\bibfield
  {journal} {\bibinfo  {journal} {Advances in Physics}\ }\textbf {\bibinfo
  {volume} {46}},\ \bibinfo {pages} {139} (\bibinfo {year}
  {1997}{\natexlab{a}})}\BibitemShut {NoStop}%
\bibitem [{\citenamefont {Salmi}\ \emph {et~al.}(1999)\citenamefont {Salmi},
  \citenamefont {Alatalo}, \citenamefont {Ala-Nissila},\ and\ \citenamefont
  {Nieminen}}]{Salmi1999}%
  \BibitemOpen
  \bibfield  {author} {\bibinfo {author} {\bibfnamefont {M.}~\bibnamefont
  {Salmi}}, \bibinfo {author} {\bibfnamefont {M.}~\bibnamefont {Alatalo}},
  \bibinfo {author} {\bibfnamefont {T.}~\bibnamefont {Ala-Nissila}},\ and\
  \bibinfo {author} {\bibfnamefont {R.}~\bibnamefont {Nieminen}},\ }\href
  {https://doi.org/https://doi.org/10.1016/S0039-6028(99)00180-6} {\bibfield
  {journal} {\bibinfo  {journal} {Surface Science}\ }\textbf {\bibinfo {volume}
  {425}},\ \bibinfo {pages} {31} (\bibinfo {year} {1999})}\BibitemShut
  {NoStop}%
\bibitem [{\citenamefont {Omi}\ \emph {et~al.}(2005)\citenamefont {Omi},
  \citenamefont {Homma}, \citenamefont {Tonchev},\ and\ \citenamefont
  {Pimpinelli}}]{Omihomma2005}%
  \BibitemOpen
  \bibfield  {author} {\bibinfo {author} {\bibfnamefont {H.}~\bibnamefont
  {Omi}}, \bibinfo {author} {\bibfnamefont {Y.}~\bibnamefont {Homma}}, \bibinfo
  {author} {\bibfnamefont {V.}~\bibnamefont {Tonchev}},\ and\ \bibinfo {author}
  {\bibfnamefont {A.}~\bibnamefont {Pimpinelli}},\ }\href
  {https://doi.org/10.1103/PhysRevLett.95.216101} {\bibfield  {journal}
  {\bibinfo  {journal} {Physical review letters}\ }\textbf {\bibinfo {volume}
  {95}},\ \bibinfo {pages} {216101} (\bibinfo {year} {2005})}\BibitemShut
  {NoStop}%
\bibitem [{\citenamefont {Ghez}\ and\ \citenamefont {Iyer}(1988)}]{Ghez1988}%
  \BibitemOpen
  \bibfield  {author} {\bibinfo {author} {\bibfnamefont {R.}~\bibnamefont
  {Ghez}}\ and\ \bibinfo {author} {\bibfnamefont {S.~S.}\ \bibnamefont
  {Iyer}},\ }\href {https://doi.org/10.1147/rd.326.0804} {\bibfield  {journal}
  {\bibinfo  {journal} {IBM Journal of Research and Development}\ }\textbf
  {\bibinfo {volume} {32}},\ \bibinfo {pages} {804} (\bibinfo {year}
  {1988})}\BibitemShut {NoStop}%
\bibitem [{\citenamefont {Michely}\ and\ \citenamefont
  {Krug}(2012)}]{Michely2012}%
  \BibitemOpen
  \bibfield  {author} {\bibinfo {author} {\bibfnamefont {T.}~\bibnamefont
  {Michely}}\ and\ \bibinfo {author} {\bibfnamefont {J.}~\bibnamefont {Krug}},\
  }\href {https://doi.org/10.1007/978-3-642-18672-1} {\emph {\bibinfo {title}
  {Islands, mounds and atoms}}},\ Vol.~\bibinfo {volume} {42}\ (\bibinfo
  {publisher} {Springer Science \& Business Media},\ \bibinfo {year}
  {2012})\BibitemShut {NoStop}%
\bibitem [{\citenamefont {Stewart}\ \emph {et~al.}(1994)\citenamefont
  {Stewart}, \citenamefont {Pohland},\ and\ \citenamefont
  {Gibson}}]{Stewart1994}%
  \BibitemOpen
  \bibfield  {author} {\bibinfo {author} {\bibfnamefont {J.}~\bibnamefont
  {Stewart}}, \bibinfo {author} {\bibfnamefont {O.}~\bibnamefont {Pohland}},\
  and\ \bibinfo {author} {\bibfnamefont {J.~M.}\ \bibnamefont {Gibson}},\
  }\href {https://doi.org/10.1103/PhysRevB.49.13848} {\bibfield  {journal}
  {\bibinfo  {journal} {Phys. Rev. B}\ }\textbf {\bibinfo {volume} {49}},\
  \bibinfo {pages} {13848} (\bibinfo {year} {1994})}\BibitemShut {NoStop}%
\bibitem [{\citenamefont {Tersoff}\ \emph {et~al.}(1995)\citenamefont
  {Tersoff}, \citenamefont {Phang}, \citenamefont {Zhang},\ and\ \citenamefont
  {Lagally}}]{Tersoff1995}%
  \BibitemOpen
  \bibfield  {author} {\bibinfo {author} {\bibfnamefont {J.}~\bibnamefont
  {Tersoff}}, \bibinfo {author} {\bibfnamefont {Y.~H.}\ \bibnamefont {Phang}},
  \bibinfo {author} {\bibfnamefont {Z.}~\bibnamefont {Zhang}},\ and\ \bibinfo
  {author} {\bibfnamefont {M.~G.}\ \bibnamefont {Lagally}},\ }\href
  {https://doi.org/10.1103/PhysRevLett.75.2730} {\bibfield  {journal} {\bibinfo
   {journal} {Phys. Rev. Lett.}\ }\textbf {\bibinfo {volume} {75}},\ \bibinfo
  {pages} {2730} (\bibinfo {year} {1995})}\BibitemShut {NoStop}%
\bibitem [{\citenamefont {Davis}(2001)}]{Davis2001}%
  \BibitemOpen
  \bibfield  {author} {\bibinfo {author} {\bibfnamefont {S.~H.}\ \bibnamefont
  {Davis}},\ }\href {https://doi.org/10.1017/CBO9780511546747} {\emph {\bibinfo
  {title} {Theory of Solidification}}},\ Cambridge Monographs on Mechanics\
  (\bibinfo  {publisher} {Cambridge University Press},\ \bibinfo {year}
  {2001})\BibitemShut {NoStop}%
\bibitem [{\citenamefont {Stoyanov}\ \emph {et~al.}(1994)\citenamefont
  {Stoyanov}, \citenamefont {Nakahara},\ and\ \citenamefont
  {Ichikawa}}]{Stoyanov1994}%
  \BibitemOpen
  \bibfield  {author} {\bibinfo {author} {\bibfnamefont {S.~S.}\ \bibnamefont
  {Stoyanov}}, \bibinfo {author} {\bibfnamefont {H.}~\bibnamefont {Nakahara}},\
  and\ \bibinfo {author} {\bibfnamefont {M.}~\bibnamefont {Ichikawa}},\ }\href
  {https://doi.org/10.1143/jjap.33.254} {\bibfield  {journal} {\bibinfo
  {journal} {Japanese Journal of Applied Physics}\ }\textbf {\bibinfo {volume}
  {33}},\ \bibinfo {pages} {254} (\bibinfo {year} {1994})}\BibitemShut
  {NoStop}%
\bibitem [{\citenamefont {Frisch}\ and\ \citenamefont
  {Verga}(2005)}]{Frisch2005}%
  \BibitemOpen
  \bibfield  {author} {\bibinfo {author} {\bibfnamefont {T.}~\bibnamefont
  {Frisch}}\ and\ \bibinfo {author} {\bibfnamefont {A.}~\bibnamefont {Verga}},\
  }\href {https://doi.org/10.1103/PhysRevLett.94.226102} {\bibfield  {journal}
  {\bibinfo  {journal} {Phys. Rev. Lett.}\ }\textbf {\bibinfo {volume} {94}},\
  \bibinfo {pages} {226102} (\bibinfo {year} {2005})}\BibitemShut {NoStop}%
\bibitem [{\citenamefont {Krug}\ \emph {et~al.}(2005)\citenamefont {Krug},
  \citenamefont {Tonchev}, \citenamefont {Stoyanov},\ and\ \citenamefont
  {Pimpinelli}}]{Krug2005_scaling}%
  \BibitemOpen
  \bibfield  {author} {\bibinfo {author} {\bibfnamefont {J.}~\bibnamefont
  {Krug}}, \bibinfo {author} {\bibfnamefont {V.}~\bibnamefont {Tonchev}},
  \bibinfo {author} {\bibfnamefont {S.}~\bibnamefont {Stoyanov}},\ and\
  \bibinfo {author} {\bibfnamefont {A.}~\bibnamefont {Pimpinelli}},\ }\href
  {https://doi.org/10.1103/PhysRevB.71.045412} {\bibfield  {journal} {\bibinfo
  {journal} {Phys. Rev. B}\ }\textbf {\bibinfo {volume} {71}},\ \bibinfo
  {pages} {045412} (\bibinfo {year} {2005})}\BibitemShut {NoStop}%
\bibitem [{\citenamefont {Liu}\ and\ \citenamefont {Metiu}(1994)}]{Liu1994}%
  \BibitemOpen
  \bibfield  {author} {\bibinfo {author} {\bibfnamefont {F.}~\bibnamefont
  {Liu}}\ and\ \bibinfo {author} {\bibfnamefont {H.}~\bibnamefont {Metiu}},\
  }\href {https://doi.org/10.1103/PhysRevE.49.2601} {\bibfield  {journal}
  {\bibinfo  {journal} {Phys. Rev. E}\ }\textbf {\bibinfo {volume} {49}},\
  \bibinfo {pages} {2601} (\bibinfo {year} {1994})}\BibitemShut {NoStop}%
\bibitem [{\citenamefont {Otto}\ \emph {et~al.}(2004)\citenamefont {Otto},
  \citenamefont {Penzler}, \citenamefont {Rätz}, \citenamefont {Rump},\ and\
  \citenamefont {Voigt}}]{Otto2004}%
  \BibitemOpen
  \bibfield  {author} {\bibinfo {author} {\bibfnamefont {F.}~\bibnamefont
  {Otto}}, \bibinfo {author} {\bibfnamefont {P.}~\bibnamefont {Penzler}},
  \bibinfo {author} {\bibfnamefont {A.}~\bibnamefont {Rätz}}, \bibinfo
  {author} {\bibfnamefont {T.}~\bibnamefont {Rump}},\ and\ \bibinfo {author}
  {\bibfnamefont {A.}~\bibnamefont {Voigt}},\ }\href
  {https://doi.org/10.1088/0951-7715/17/2/006} {\bibfield  {journal} {\bibinfo
  {journal} {Nonlinearity}\ }\textbf {\bibinfo {volume} {17}},\ \bibinfo
  {pages} {477} (\bibinfo {year} {2004})}\BibitemShut {NoStop}%
\bibitem [{\citenamefont {Torabi}\ \emph {et~al.}(2009)\citenamefont {Torabi},
  \citenamefont {Lowengrub}, \citenamefont {Voigt},\ and\ \citenamefont
  {Wise}}]{Torabi2009}%
  \BibitemOpen
  \bibfield  {author} {\bibinfo {author} {\bibfnamefont {S.}~\bibnamefont
  {Torabi}}, \bibinfo {author} {\bibfnamefont {J.}~\bibnamefont {Lowengrub}},
  \bibinfo {author} {\bibfnamefont {A.}~\bibnamefont {Voigt}},\ and\ \bibinfo
  {author} {\bibfnamefont {S.}~\bibnamefont {Wise}},\ }\href
  {https://doi.org/10.1098/rspa.2008.0385} {\bibfield  {journal} {\bibinfo
  {journal} {Proceedings of the Royal Society A: Mathematical, Physical and
  Engineering Sciences}\ }\textbf {\bibinfo {volume} {465}},\ \bibinfo {pages}
  {1337} (\bibinfo {year} {2009})}\BibitemShut {NoStop}%
\bibitem [{\citenamefont {Hu}\ \emph {et~al.}(2012)\citenamefont {Hu},
  \citenamefont {Lowengrub}, \citenamefont {Wise},\ and\ \citenamefont
  {Voigt}}]{Hu2012}%
  \BibitemOpen
  \bibfield  {author} {\bibinfo {author} {\bibfnamefont {Z.}~\bibnamefont
  {Hu}}, \bibinfo {author} {\bibfnamefont {J.~S.}\ \bibnamefont {Lowengrub}},
  \bibinfo {author} {\bibfnamefont {S.~M.}\ \bibnamefont {Wise}},\ and\
  \bibinfo {author} {\bibfnamefont {A.}~\bibnamefont {Voigt}},\ }\href
  {https://doi.org/https://doi.org/10.1016/j.physd.2011.09.004} {\bibfield
  {journal} {\bibinfo  {journal} {Physica D: Nonlinear Phenomena}\ }\textbf
  {\bibinfo {volume} {241}},\ \bibinfo {pages} {77} (\bibinfo {year}
  {2012})}\BibitemShut {NoStop}%
\bibitem [{\citenamefont {Rätz}\ and\ \citenamefont {Voigt}(2004)}]{Ratz2004}%
  \BibitemOpen
  \bibfield  {author} {\bibinfo {author} {\bibfnamefont {A.}~\bibnamefont
  {Rätz}}\ and\ \bibinfo {author} {\bibfnamefont {A.}~\bibnamefont {Voigt}},\
  }\href {https://doi.org/https://doi.org/10.1016/j.jcrysgro.2004.02.075}
  {\bibfield  {journal} {\bibinfo  {journal} {Journal of Crystal Growth}\
  }\textbf {\bibinfo {volume} {266}},\ \bibinfo {pages} {278} (\bibinfo {year}
  {2004})},\ \bibinfo {note} {proceedings of the Fourth International Workshop
  on Modeling in Crystal Growth}\BibitemShut {NoStop}%
\bibitem [{\citenamefont {Yu}\ \emph {et~al.}(2011)\citenamefont {Yu},
  \citenamefont {Voigt}, \citenamefont {Guo},\ and\ \citenamefont
  {Liu}}]{Yu2011}%
  \BibitemOpen
  \bibfield  {author} {\bibinfo {author} {\bibfnamefont {Y.-M.}\ \bibnamefont
  {Yu}}, \bibinfo {author} {\bibfnamefont {A.}~\bibnamefont {Voigt}}, \bibinfo
  {author} {\bibfnamefont {X.}~\bibnamefont {Guo}},\ and\ \bibinfo {author}
  {\bibfnamefont {Y.}~\bibnamefont {Liu}},\ }\href
  {https://doi.org/10.1063/1.3666781} {\bibfield  {journal} {\bibinfo
  {journal} {Applied Physics Letters}\ }\textbf {\bibinfo {volume} {99}},\
  \bibinfo {pages} {263106} (\bibinfo {year} {2011})}\BibitemShut {NoStop}%
\bibitem [{Jul()}]{Julia}%
  \BibitemOpen
  \href@noop {} {\bibinfo {title} {{T}he {J}ulia {P}rogramming {L}anguage}},\
  \bibinfo {howpublished} {\url{https://julialang.org}},\ \bibinfo {note}
  {{A}ccessed: 2021-04-28}\BibitemShut {NoStop}%
\bibitem [{Sun()}]{Sundials}%
  \BibitemOpen
  \href@noop {} {\bibinfo {title} {{CVODE} solver description}},\ \bibinfo
  {howpublished} {\url{https://computing.llnl.gov/projects/sundials/cvode}},\
  \bibinfo {note} {{A}ccessed: 2021-04-28}\BibitemShut {NoStop}%
\bibitem [{\citenamefont {Sato}\ and\ \citenamefont {Uwaha}(1999)}]{Sato1999}%
  \BibitemOpen
  \bibfield  {author} {\bibinfo {author} {\bibfnamefont {M.}~\bibnamefont
  {Sato}}\ and\ \bibinfo {author} {\bibfnamefont {M.}~\bibnamefont {Uwaha}},\
  }\href {https://doi.org/https://doi.org/10.1016/S0039-6028(99)00932-2}
  {\bibfield  {journal} {\bibinfo  {journal} {Surface Science}\ }\textbf
  {\bibinfo {volume} {442}},\ \bibinfo {pages} {318} (\bibinfo {year}
  {1999})}\BibitemShut {NoStop}%
\bibitem [{\citenamefont {Homma}\ and\ \citenamefont
  {Aizawa}(2000)}]{Homma2000}%
  \BibitemOpen
  \bibfield  {author} {\bibinfo {author} {\bibfnamefont {Y.}~\bibnamefont
  {Homma}}\ and\ \bibinfo {author} {\bibfnamefont {N.}~\bibnamefont {Aizawa}},\
  }\href {https://doi.org/10.1103/PhysRevB.62.8323} {\bibfield  {journal}
  {\bibinfo  {journal} {Phys. Rev. B}\ }\textbf {\bibinfo {volume} {62}},\
  \bibinfo {pages} {8323} (\bibinfo {year} {2000})}\BibitemShut {NoStop}%
\bibitem [{\citenamefont {Toktarbaiuly}\ \emph {et~al.}(2018)\citenamefont
  {Toktarbaiuly}, \citenamefont {Usov}, \citenamefont {\'O~Coile\'ain},
  \citenamefont {Siewierska}, \citenamefont {Krasnikov}, \citenamefont
  {Norton}, \citenamefont {Bozhko}, \citenamefont {Semenov}, \citenamefont
  {Chaika}, \citenamefont {Murphy}, \citenamefont {L\"ubben}, \citenamefont
  {Krzy\ifmmode~\dot{z}\else \.{z}\fi{}ewski}, \citenamefont
  {Za\l{}uska-Kotur}, \citenamefont {Krasteva}, \citenamefont {Popova},
  \citenamefont {Tonchev},\ and\ \citenamefont {Shvets}}]{Toktarbaiuly2018}%
  \BibitemOpen
  \bibfield  {author} {\bibinfo {author} {\bibfnamefont {O.}~\bibnamefont
  {Toktarbaiuly}}, \bibinfo {author} {\bibfnamefont {V.}~\bibnamefont {Usov}},
  \bibinfo {author} {\bibfnamefont {C.}~\bibnamefont {\'O~Coile\'ain}},
  \bibinfo {author} {\bibfnamefont {K.}~\bibnamefont {Siewierska}}, \bibinfo
  {author} {\bibfnamefont {S.}~\bibnamefont {Krasnikov}}, \bibinfo {author}
  {\bibfnamefont {E.}~\bibnamefont {Norton}}, \bibinfo {author} {\bibfnamefont
  {S.~I.}\ \bibnamefont {Bozhko}}, \bibinfo {author} {\bibfnamefont {V.~N.}\
  \bibnamefont {Semenov}}, \bibinfo {author} {\bibfnamefont {A.~N.}\
  \bibnamefont {Chaika}}, \bibinfo {author} {\bibfnamefont {B.~E.}\
  \bibnamefont {Murphy}}, \bibinfo {author} {\bibfnamefont {O.}~\bibnamefont
  {L\"ubben}}, \bibinfo {author} {\bibfnamefont {F.}~\bibnamefont
  {Krzy\ifmmode~\dot{z}\else \.{z}\fi{}ewski}}, \bibinfo {author}
  {\bibfnamefont {M.~A.}\ \bibnamefont {Za\l{}uska-Kotur}}, \bibinfo {author}
  {\bibfnamefont {A.}~\bibnamefont {Krasteva}}, \bibinfo {author}
  {\bibfnamefont {H.}~\bibnamefont {Popova}}, \bibinfo {author} {\bibfnamefont
  {V.}~\bibnamefont {Tonchev}},\ and\ \bibinfo {author} {\bibfnamefont {I.~V.}\
  \bibnamefont {Shvets}},\ }\href {https://doi.org/10.1103/PhysRevB.97.035436}
  {\bibfield  {journal} {\bibinfo  {journal} {Phys. Rev. B}\ }\textbf {\bibinfo
  {volume} {97}},\ \bibinfo {pages} {035436} (\bibinfo {year}
  {2018})}\BibitemShut {NoStop}%
\bibitem [{\citenamefont {Slanina}\ \emph {et~al.}(2005)\citenamefont
  {Slanina}, \citenamefont {Krug},\ and\ \citenamefont {Kotrla}}]{Slanina2005}%
  \BibitemOpen
  \bibfield  {author} {\bibinfo {author} {\bibfnamefont {F.}~\bibnamefont
  {Slanina}}, \bibinfo {author} {\bibfnamefont {J.}~\bibnamefont {Krug}},\ and\
  \bibinfo {author} {\bibfnamefont {M.}~\bibnamefont {Kotrla}},\ }\href
  {https://doi.org/10.1103/PhysRevE.71.041605} {\bibfield  {journal} {\bibinfo
  {journal} {Phys. Rev. E}\ }\textbf {\bibinfo {volume} {71}},\ \bibinfo
  {pages} {041605} (\bibinfo {year} {2005})}\BibitemShut {NoStop}%
\bibitem [{\citenamefont {Krzyżewski}\ \emph {et~al.}(2017)\citenamefont
  {Krzyżewski}, \citenamefont {Załuska-Kotur}, \citenamefont {Krasteva},
  \citenamefont {Popova},\ and\ \citenamefont {Tonchev}}]{Krzyzewski2017}%
  \BibitemOpen
  \bibfield  {author} {\bibinfo {author} {\bibfnamefont {F.}~\bibnamefont
  {Krzyżewski}}, \bibinfo {author} {\bibfnamefont {M.}~\bibnamefont
  {Załuska-Kotur}}, \bibinfo {author} {\bibfnamefont {A.}~\bibnamefont
  {Krasteva}}, \bibinfo {author} {\bibfnamefont {H.}~\bibnamefont {Popova}},\
  and\ \bibinfo {author} {\bibfnamefont {V.}~\bibnamefont {Tonchev}},\ }\href
  {https://doi.org/https://doi.org/10.1016/j.jcrysgro.2016.11.121} {\bibfield
  {journal} {\bibinfo  {journal} {Journal of Crystal Growth}\ }\textbf
  {\bibinfo {volume} {474}},\ \bibinfo {pages} {135} (\bibinfo {year}
  {2017})},\ \bibinfo {note} {the 8th International Workshop on Modeling in
  Crystal Growth}\BibitemShut {NoStop}%
\bibitem [{\citenamefont {Tonchev}(2012)}]{Tonchev2012}%
  \BibitemOpen
  \bibfield  {author} {\bibinfo {author} {\bibfnamefont {V.}~\bibnamefont
  {Tonchev}},\ }\href
  {http://www.bcc.bas.bg/BCC_Volumes/Volume_44_Special_2012/Volume_44_Special_2012_PDF/BCC-44-Special_Issue-17.pdf}
  {\bibfield  {journal} {\bibinfo  {journal} {Bulgarian Chemical
  Communications}\ }\textbf {\bibinfo {volume} {44}} (\bibinfo {year}
  {2012})}\BibitemShut {NoStop}%
\bibitem [{\citenamefont {Ichimiya}\ \emph {et~al.}(2000)\citenamefont
  {Ichimiya}, \citenamefont {Hayashi}, \citenamefont {Williams}, \citenamefont
  {Einstein}, \citenamefont {Uwaha},\ and\ \citenamefont
  {Watanabe}}]{Ichimiya2000}%
  \BibitemOpen
  \bibfield  {author} {\bibinfo {author} {\bibfnamefont {A.}~\bibnamefont
  {Ichimiya}}, \bibinfo {author} {\bibfnamefont {K.}~\bibnamefont {Hayashi}},
  \bibinfo {author} {\bibfnamefont {E.~D.}\ \bibnamefont {Williams}}, \bibinfo
  {author} {\bibfnamefont {T.~L.}\ \bibnamefont {Einstein}}, \bibinfo {author}
  {\bibfnamefont {M.}~\bibnamefont {Uwaha}},\ and\ \bibinfo {author}
  {\bibfnamefont {K.}~\bibnamefont {Watanabe}},\ }\href
  {https://doi.org/10.1103/PhysRevLett.84.3662} {\bibfield  {journal} {\bibinfo
   {journal} {Phys. Rev. Lett.}\ }\textbf {\bibinfo {volume} {84}},\ \bibinfo
  {pages} {3662} (\bibinfo {year} {2000})}\BibitemShut {NoStop}%
\bibitem [{\citenamefont {Yang}\ and\ \citenamefont
  {Williams}(1994)}]{Yang1994}%
  \BibitemOpen
  \bibfield  {author} {\bibinfo {author} {\bibfnamefont {Y.-N.}\ \bibnamefont
  {Yang}}\ and\ \bibinfo {author} {\bibfnamefont {E.~D.}\ \bibnamefont
  {Williams}},\ }\href {https://doi.org/10.1103/PhysRevLett.72.1862} {\bibfield
   {journal} {\bibinfo  {journal} {Phys. Rev. Lett.}\ }\textbf {\bibinfo
  {volume} {72}},\ \bibinfo {pages} {1862} (\bibinfo {year}
  {1994})}\BibitemShut {NoStop}%
\bibitem [{\citenamefont {Johnson}\ \emph {et~al.}(1996)\citenamefont
  {Johnson}, \citenamefont {Leung}, \citenamefont {Birch}, \citenamefont
  {Orr},\ and\ \citenamefont {Tersoff}}]{Johnson1996}%
  \BibitemOpen
  \bibfield  {author} {\bibinfo {author} {\bibfnamefont {M.}~\bibnamefont
  {Johnson}}, \bibinfo {author} {\bibfnamefont {K.}~\bibnamefont {Leung}},
  \bibinfo {author} {\bibfnamefont {A.}~\bibnamefont {Birch}}, \bibinfo
  {author} {\bibfnamefont {B.}~\bibnamefont {Orr}},\ and\ \bibinfo {author}
  {\bibfnamefont {J.}~\bibnamefont {Tersoff}},\ }\href
  {https://doi.org/https://doi.org/10.1016/0039-6028(95)01110-2} {\bibfield
  {journal} {\bibinfo  {journal} {Surface Science}\ }\textbf {\bibinfo {volume}
  {350}},\ \bibinfo {pages} {254} (\bibinfo {year} {1996})}\BibitemShut
  {NoStop}%
\bibitem [{\citenamefont {Schelling}\ \emph {et~al.}(2000)\citenamefont
  {Schelling}, \citenamefont {Springholz},\ and\ \citenamefont
  {Schäffler}}]{Schelling2000}%
  \BibitemOpen
  \bibfield  {author} {\bibinfo {author} {\bibfnamefont {C.}~\bibnamefont
  {Schelling}}, \bibinfo {author} {\bibfnamefont {G.}~\bibnamefont
  {Springholz}},\ and\ \bibinfo {author} {\bibfnamefont {F.}~\bibnamefont
  {Schäffler}},\ }\href
  {https://doi.org/https://doi.org/10.1016/S0040-6090(00)00823-3} {\bibfield
  {journal} {\bibinfo  {journal} {Thin Solid Films}\ }\textbf {\bibinfo
  {volume} {369}},\ \bibinfo {pages} {1 } (\bibinfo {year} {2000})}\BibitemShut
  {NoStop}%
\bibitem [{\citenamefont {Ishizaki}\ \emph {et~al.}(1996)\citenamefont
  {Ishizaki}, \citenamefont {Ohkuri},\ and\ \citenamefont
  {Fukui}}]{Ishizaki1996}%
  \BibitemOpen
  \bibfield  {author} {\bibinfo {author} {\bibfnamefont {J.}~\bibnamefont
  {Ishizaki}}, \bibinfo {author} {\bibfnamefont {K.}~\bibnamefont {Ohkuri}},\
  and\ \bibinfo {author} {\bibfnamefont {T.}~\bibnamefont {Fukui}},\ }\href
  {https://doi.org/10.1143/jjap.35.1280} {\bibfield  {journal} {\bibinfo
  {journal} {Japanese Journal of Applied Physics}\ }\textbf {\bibinfo {volume}
  {35}},\ \bibinfo {pages} {1280} (\bibinfo {year} {1996})}\BibitemShut
  {NoStop}%
\bibitem [{\citenamefont {Pimpinelli}\ \emph {et~al.}(2002)\citenamefont
  {Pimpinelli}, \citenamefont {Tonchev}, \citenamefont {Videcoq},\ and\
  \citenamefont {Vladimirova}}]{PTVV2002}%
  \BibitemOpen
  \bibfield  {author} {\bibinfo {author} {\bibfnamefont {A.}~\bibnamefont
  {Pimpinelli}}, \bibinfo {author} {\bibfnamefont {V.}~\bibnamefont {Tonchev}},
  \bibinfo {author} {\bibfnamefont {A.}~\bibnamefont {Videcoq}},\ and\ \bibinfo
  {author} {\bibfnamefont {M.}~\bibnamefont {Vladimirova}},\ }\href
  {https://doi.org/10.1103/PhysRevLett.88.206103} {\bibfield  {journal}
  {\bibinfo  {journal} {Physical review letters}\ }\textbf {\bibinfo {volume}
  {88}},\ \bibinfo {pages} {206103} (\bibinfo {year} {2002})}\BibitemShut
  {NoStop}%
\bibitem [{\citenamefont {Vladimirova}\ \emph {et~al.}(2001)\citenamefont
  {Vladimirova}, \citenamefont {De~Vita},\ and\ \citenamefont
  {Pimpinelli}}]{Vladimirova2001}%
  \BibitemOpen
  \bibfield  {author} {\bibinfo {author} {\bibfnamefont {M.}~\bibnamefont
  {Vladimirova}}, \bibinfo {author} {\bibfnamefont {A.}~\bibnamefont
  {De~Vita}},\ and\ \bibinfo {author} {\bibfnamefont {A.}~\bibnamefont
  {Pimpinelli}},\ }\href {https://doi.org/10.1103/PhysRevB.64.245420}
  {\bibfield  {journal} {\bibinfo  {journal} {Phys. Rev. B}\ }\textbf {\bibinfo
  {volume} {64}},\ \bibinfo {pages} {245420} (\bibinfo {year}
  {2001})}\BibitemShut {NoStop}%
\bibitem [{\citenamefont {{Popkov}}\ and\ \citenamefont
  {{Krug}}(2005)}]{Popkov_Krug_moving_bunches}%
  \BibitemOpen
  \bibfield  {author} {\bibinfo {author} {\bibfnamefont {V.}~\bibnamefont
  {{Popkov}}}\ and\ \bibinfo {author} {\bibfnamefont {J.}~\bibnamefont
  {{Krug}}},\ }\href {https://doi.org/10.1209/epl/i2005-10335-4} {\bibfield
  {journal} {\bibinfo  {journal} {EPL (Europhysics Letters)}\ }\textbf
  {\bibinfo {volume} {72}},\ \bibinfo {pages} {1025} (\bibinfo {year}
  {2005})}\BibitemShut {NoStop}%
\bibitem [{\citenamefont {Krug}(1997{\natexlab{b}})}]{Krug_homogenization}%
  \BibitemOpen
  \bibfield  {author} {\bibinfo {author} {\bibfnamefont {J.}~\bibnamefont
  {Krug}},\ }in\ \href {https://doi.org/10.1142/3501} {\emph {\bibinfo
  {booktitle} {Dynamics of Fluctuating Interfaces and Related Phenomena}}},\
  \bibinfo {editor} {edited by\ \bibinfo {editor} {\bibfnamefont
  {D.}~\bibnamefont {Kim}}, \bibinfo {editor} {\bibfnamefont {H.}~\bibnamefont
  {Park}},\ and\ \bibinfo {editor} {\bibfnamefont {B.}~\bibnamefont {Kahng}}}\
  (\bibinfo  {publisher} {World Scientific},\ \bibinfo {year}
  {1997})\BibitemShut {NoStop}%
\bibitem [{\citenamefont {Xiang}(2002)}]{Xiang2002}%
  \BibitemOpen
  \bibfield  {author} {\bibinfo {author} {\bibfnamefont {Y.}~\bibnamefont
  {Xiang}},\ }\href {https://doi.org/10.1137/S003613990139828X} {\bibfield
  {journal} {\bibinfo  {journal} {SIAM Journal on Applied Mathematics}\
  }\textbf {\bibinfo {volume} {63}},\ \bibinfo {pages} {241} (\bibinfo {year}
  {2002})}\BibitemShut {NoStop}%
\bibitem [{\citenamefont {Margetis}\ \emph {et~al.}(2005)\citenamefont
  {Margetis}, \citenamefont {Aziz},\ and\ \citenamefont
  {Stone}}]{Margetis2005}%
  \BibitemOpen
  \bibfield  {author} {\bibinfo {author} {\bibfnamefont {D.}~\bibnamefont
  {Margetis}}, \bibinfo {author} {\bibfnamefont {M.~J.}\ \bibnamefont {Aziz}},\
  and\ \bibinfo {author} {\bibfnamefont {H.~A.}\ \bibnamefont {Stone}},\ }\href
  {https://doi.org/10.1103/PhysRevB.71.165432} {\bibfield  {journal} {\bibinfo
  {journal} {Phys. Rev. B}\ }\textbf {\bibinfo {volume} {71}},\ \bibinfo
  {pages} {165432} (\bibinfo {year} {2005})}\BibitemShut {NoStop}%
\bibitem [{\citenamefont {Warming}\ and\ \citenamefont
  {Hyett}(1974)}]{Warming1974}%
  \BibitemOpen
  \bibfield  {author} {\bibinfo {author} {\bibfnamefont {R.}~\bibnamefont
  {Warming}}\ and\ \bibinfo {author} {\bibfnamefont {B.}~\bibnamefont
  {Hyett}},\ }\href
  {https://doi.org/https://doi.org/10.1016/0021-9991(74)90011-4} {\bibfield
  {journal} {\bibinfo  {journal} {Journal of Computational Physics}\ }\textbf
  {\bibinfo {volume} {14}},\ \bibinfo {pages} {159} (\bibinfo {year}
  {1974})}\BibitemShut {NoStop}%
\bibitem [{\citenamefont {Margetis}\ and\ \citenamefont
  {Kohn}(2006)}]{Margetis2006}%
  \BibitemOpen
  \bibfield  {author} {\bibinfo {author} {\bibfnamefont {D.}~\bibnamefont
  {Margetis}}\ and\ \bibinfo {author} {\bibfnamefont {R.~V.}\ \bibnamefont
  {Kohn}},\ }\href {https://doi.org/10.1137/06065297X} {\bibfield  {journal}
  {\bibinfo  {journal} {Multiscale Modeling \& Simulation}\ }\textbf {\bibinfo
  {volume} {5}},\ \bibinfo {pages} {729} (\bibinfo {year} {2006})}\BibitemShut
  {NoStop}%
\bibitem [{\citenamefont {Schwoebel}\ and\ \citenamefont
  {Shipsey}(1966)}]{Schwoebel1966}%
  \BibitemOpen
  \bibfield  {author} {\bibinfo {author} {\bibfnamefont {R.~L.}\ \bibnamefont
  {Schwoebel}}\ and\ \bibinfo {author} {\bibfnamefont {E.~J.}\ \bibnamefont
  {Shipsey}},\ }\href {https://doi.org/10.1063/1.1707904} {\bibfield  {journal}
  {\bibinfo  {journal} {Journal of Applied Physics}\ }\textbf {\bibinfo
  {volume} {37}},\ \bibinfo {pages} {3682} (\bibinfo {year}
  {1966})}\BibitemShut {NoStop}%
\bibitem [{\citenamefont {Guin}(2018)}]{Guin_thesis}%
  \BibitemOpen
  \bibfield  {author} {\bibinfo {author} {\bibfnamefont {L.}~\bibnamefont
  {Guin}},\ }\emph {\bibinfo {title} {{Electromechanical couplings and growth
  instabilities in semiconductors}}},\ \href
  {https://tel.archives-ouvertes.fr/tel-02124459} {\bibinfo {type} {Ph{D}
  thesis}},\ \bibinfo  {school} {{Universit{\'e} Paris-Saclay}} (\bibinfo
  {year} {2018})\BibitemShut {NoStop}%
\bibitem [{\citenamefont {Stoyanov}(2000)}]{Stoyanov2000}%
  \BibitemOpen
  \bibfield  {author} {\bibinfo {author} {\bibfnamefont {S.}~\bibnamefont
  {Stoyanov}},\ }\href
  {https://doi.org/https://doi.org/10.1016/S0039-6028(00)00701-9} {\bibfield
  {journal} {\bibinfo  {journal} {Surface Science}\ }\textbf {\bibinfo {volume}
  {464}},\ \bibinfo {pages} {L715} (\bibinfo {year} {2000})}\BibitemShut
  {NoStop}%
\end{thebibliography}%

\appendix
\section{Discretization of \eqref{eq:ale} using the Galerkin method}\label{app:galerkin}
We multiply \eqref{eq:ale} by a weight function $\phi$ and integrate:
\begin{align}
\begin{split}
	\Pc\int_0^1\dt\rnb \phi \ud u =\ &\frac{1}{\ol{s}_n^2}\int_0^1(\duu\rnb)\phi \ud u \\
	&+ \Pc\frac{\dot{\ol{x}}_n}{\ol{s}_n}\int_0^1(\du\rnb)\phi \ud du \\
	&+ \Pc\frac{\dot{\ol{s}}_n}{\ol{s}_n}\int_0^1(u\hspace{1pt}\du\rnb) \phi \ud u \\
	&+ \Fb\int_0^1\phi \ud u.
\end{split}
\end{align}
Next, integrating by parts the term with the double derivative, we get:
\begin{align}
\begin{split}
&\Pc\int_0^1\dt\rnb \phi \ud u = \frac{1}{\ol{s}_n^2}\Big[ (\du\rnb)\phi \Big]_0^1 \\
&\quad- \frac{1}{\ol{s}_n^2}\int_0^1(\du\rnb)\phi' \ud u + \Pc\frac{\dot{\ol{x}}_n}{\ol{s}_n}\int_0^1(\du\rnb)\phi \ud u \\
&\quad\quad+ \Pc\frac{\dot{\ol{s}}_n}{\ol{s}_n}\int_0^1(u\hspace{1pt}\du\rnb) \phi \ud u + \Fb\int_0^1\phi \ud u.
\end{split}
\end{align}
Finally, introducing some shape functions $\varphi_i$, we write $\rnb(\ol{t},u)\!=\!\sum_i q_i^n(\ol{t})\varphi_i(u)$ and substitute $\phi$ for an arbitrary $\varphi_j$ to obtain the following system:
\begin{align}
\begin{split}
\Pc\hspace{1pt}\mathbf{M}\mathbf{\dot{q}}^n =\ & \mathbf{A}\!^n - \frac{1}{\ol{s}_n^2}\mathbf{D}^{(2)}\mathbf{q}^n + \Pc\frac{\dot{\ol{x}}_n}{\ol{s}_n}\mathbf{D}^{(1)}\mathbf{q}^n \\
&+ \Pc\frac{\dot{\ol{s}}_n}{\ol{s}_n}\mathbf{D}^{(u)}\mathbf{q}^n + \Fb\hspace{0.5pt}\mathbf{B},
\end{split}
\end{align}
at the $n$th terrace $(n\in\mathbb{N}*)$, where, from the boundary conditions \eqref{BC1d} and \eqref{BC2d},
\begin{align}
\begin{split}
	A_i^n =& \frac{1}{\ol{s}_n^2}\Big[(\du\rnb)\varphi_i\Big]_0^1 \\
	   =& \frac{1}{\ol{s}_n}\Big(-\ol{J}_{n+1}^- - \rnb^-\dot{\ol{x}}_{n+1}\Big)\varphi_i(1) \\
	   &- \frac{1}{\ol{s}_n}\Big(\ol{J}_n^+ - \rnb^+\dot{\ol{x}}_{n}\Big)\varphi_i(0),
\end{split}
\end{align} 
and
\begin{empheq}[left=\empheqlbrace]{align}
\begin{aligned}
	M_{ij} &= \int_0^1 \varphi_i(u)\varphi_j(u) \ud u, \\
	D^{(2)}_{ij} &= \int_0^1 \varphi_i^\prime(u)\varphi_j^\prime(u) \ud u, \\
	D^{(1)}_{ij} &= \int_0^1 \varphi_i(u)\varphi_j^\prime(u) \ud u, \\
	D^{(u)}_{ij} &= \int_0^1 u\varphi_i(u)\varphi_j^\prime(u) \ud u, \\
	B_i &= \int_0^1 \varphi_i(u) \ud u.
\end{aligned}
\end{empheq}

\section{Solutions of equation \eqref{rho_sol}}\label{app:rho_sol}
The functions  $\psi_n$, $\varphi_n$, and $c_n$ introduced in \eqref{rho_sol} are given by
\begin{empheq}[left=\empheqlbrace]{align}
\begin{aligned}
\psi_n(\ol{t},u) &= \frac{\exp(-\chi_a \Pc \dot{\ol{x}}_n \ol{s}_n u) - 1}{\exp(-\chi_a \Pc \dot{\ol{x}}_n \ol{s}_n) - 1}, \\
\varphi_n(\ol{t},u) &= 1 - \psi_n(\ol{t},u), \\
c_n(\ol{t},u) &= \frac{1}{\Theta}\frac{\ol{s}_n}{\chi_a \dot{\ol{x}}_n}(\psi_n(\ol{t},u) - u).
\end{aligned}
\end{empheq}
Further, letting $R_n = T_n^{(0)} T_n^{(1)} - T_n^{(2)} T_n^{(3)}$, the boundary values can be expressed as
\begin{empheq}[left=\empheqlbrace]{align}
\begin{aligned}
\rnt^+ &= \frac{1}{R_n}\Big( T_n^{(3)} T_n^{(4)} - T_n^{(0)} T_n^{(5)} \Big), \\
\rnt^- &= \frac{1}{R_n}\Big( T_n^{(2)} T_n^{(5)} - T_n^{(1)} T_n^{(4)} \Big).
\end{aligned}
\end{empheq}
where
\begin{empheq}[left=\empheqlbrace]{align}
\begin{aligned}
T_n^{(0)} &= \frac{1}{\ol{s}_n}\psi'_n(0),\\
\mbox{}&\\[-1em]
T_n^{(1)} &= \frac{1}{\ol{s}_n}\varphi'_n(1),\\
T_n^{(2)} &= \frac{1}{\ol{s}_n}\varphi'_n(0) - \frac{\ka S}{C_1} + \chi_a \Pc\dot{\ol{x}}_n,\\
T_n^{(3)} &= \frac{1}{\ol{s}_n}\psi'_n(1) + \frac{\ka S}{C_2}  + \chi_a \Pc\dot{\ol{x}}_{n+1},\\
T_n^{(4)} &= \frac{1}{\ol{s}_n}c'_n(0) - \frac{\ka S}{C_1}(\mathfrak{f}_n-1)     - \frac{\chi_c S}{C_1} \Pc \dot{\ol{x}}_n,\\
T_n^{(5)} &= \frac{1}{\ol{s}_n}c'_n(1) + \frac{\ka S}{C_2}(\mathfrak{f}_{n+1}-1) - \frac{\chi_c}{C_2}   \Pc \dot{\ol{x}}_{n+1}.
\end{aligned}
\end{empheq}

\section{Approximate solutions of \eqref{eq:traveling-wave}}\label{app:pade_coeff}
The coefficients $a_1, a_2, b_2$, and $b_3$ in \eqref{eq:pade_approximant} are determined as follows. We substitute \eqref{eq:pade_approximant} in \eqref{eq:traveling-wave}, Taylor-expand in $\xi$, and look at the four lowest orders, which yields a nonlinear system of four equations with the four coefficients as unknowns. 

As this system is impossible to solve analytically, we instead solve an approximated version. We proceed by first solving the system numerically for a wide range of model parameters to determine the dominant scaling of the unknown coefficients with $\eps$, and find
\begin{empheq}[left=\empheqlbrace]{align}
\begin{aligned}
&& a_1 &\sim \eps^{-5/3},\ && a_2 \sim \eps^{-8/3}, \\ 
&& b_2 &\sim \eps^{-2},\ && b_3 \sim \eps^{-3}.
\end{aligned}
\end{empheq}
Next, using these scaling relations and $M\!=\ell_{min}^{-1}\!\sim \eps^{-2/3}$, we approximate each equation of the system by its two leading contributions in $\eps$:
\begin{empheq}[left=\empheqlbrace]{align}
\begin{aligned}
&-v(1+M) + \frac{12 K_2}{M}\Sigma_2\eps^2 = 0, \\[0.5em]
&- 16 K_2 (-3 a_1\Sigma_2 + M \Sigma_1 (a_2 + 4 M b_2))\eps^2 = 0, \\[0.5em]
&-120 K_2 M^2 a_1^2\Sigma_2\eps^2 + M^5 \Sigma_1 v \\
&\quad+ 24 K_2 M^3 (a_1 \Sigma_1 (3 a_2 + 13 M b_2)\eps^2 \\
&\quad+ M^2 b_3(8 a_2 + 13 M b_2)) \eps^2 = 0,\\[0.5em]
&- M^3 \Sigma_2 v + 8 K_2( 30 a_1^3 \Sigma_2 \\
&\quad- 5 M a_1^2 \Sigma_1 (5 a_2 + 23 M b_2) \\
&\quad- 2 M^3 a_1 b_3 (47 a_2 + 77 M b_2) \\
&\quad+ M^2(-4 a_2^2(a_2 - M b_2) \\
&\quad\quad+ 50 M^2 a_2 b_2^2 + 3 M^3(14 b_2^3 - 13 b_3^2)))\eps^2 = 0.
\end{aligned}
\end{empheq}
where
\begin{empheq}[left=\empheqlbrace]{align}
\begin{aligned}
\Sigma_1 &= a_2 + M b_2, \\
\Sigma_2 &= a_1 \Sigma_1 + M^2 b_3.
\end{aligned}
\end{empheq}
This approximate system can now be solved analytically and yields \eqref{eq:pade_coeff} at leading order in $\eps$.

\end{document}